# Страсти по гравитационным волнам

**З.К. Силагадзе**

**Институт Ядерной Физики им. Г.И. Будкера и Новосибирский Государственный Университет**

**Недавно коллаборация LIGO зарегистрировала уже третью [1] (и даже четвертую [2]) гравитационную волну. Без всякого сомнения, открытие гравитационных волн – это самое важное событие в физике последнего десятилетия наряду с обнаружением бозона Хиггса [3]. Это эпохальное открытие является еще одним подтверждением общей теории относительности, на основе которой Альберт Эйнштейн и предсказал существование гравитационных волн в 1916 году.**

Однако, по справедливости, историю гравитационных волн стоит начать с 1905 года [4], когда Анри Пуанкаре (на фотографии) опубликовал работу «О динамике электрона» (хотя еще раньше Хевисайд в 1893 году и Лоренц в 1900 году тоже предположили существование гравитационных волн – см, статью E. Amaldi, The search for gravitational waves в книге Proceedings of the NATO Advanced Study Institute on Cosmic Gamma Rays and Cosmic Neutrinos Erice, Italy, 1988). В этой работе Пуанкаре сформулировал основы теории, которая потом будет известна под именем «специальная теория относительности». К сожалению, эта опередившая свое время работа не оказала значительного влияния [5] на развитие специальной теории относительности в годы становления этой теории, так как была математически более сложной по сравнению со знаменитой работой Эйнштейна «К электродинамике движущихся тел», вышедшей почти сразу после работы Пуанкаре.

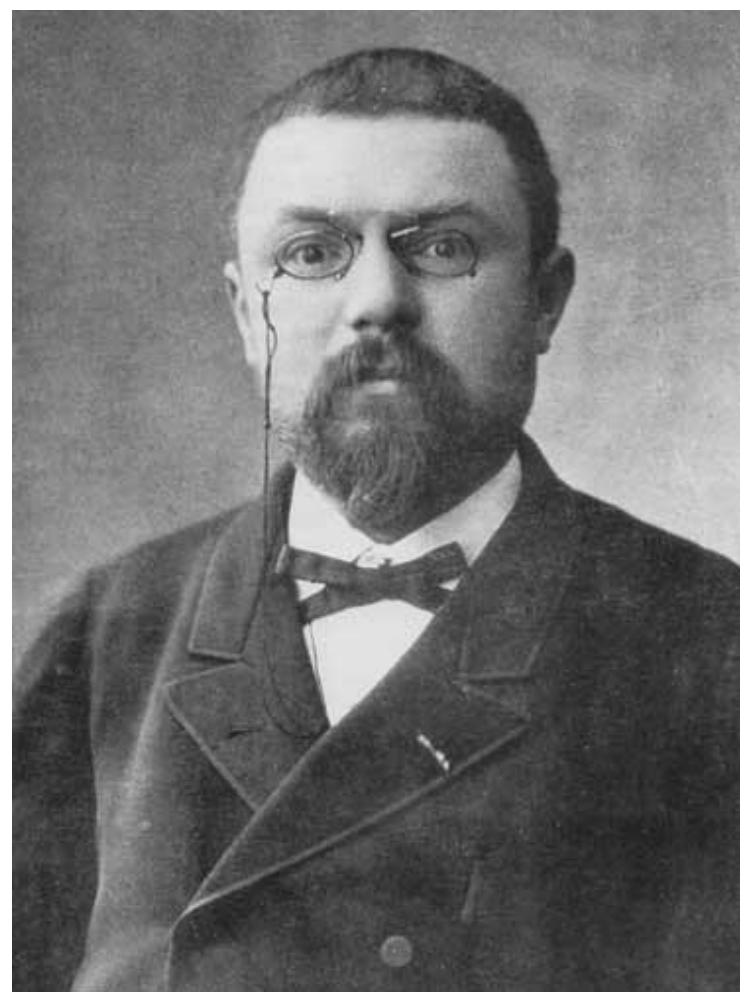

Пуанкаре правильно понимал, что симметрия относительно группы Лоренца является свойством не только электромагнитных явлений, но и вообще всех явлений природы. Поэтому в своей работе он впервые попытался сформулировать релятивистскую теорию гравитации. Из этой теории вытекало, что гравитационные взаимодействия распространяются с конечной скоростью, а именно со скоростью света, и что должны существовать гравитационные волны, аналогичные электромагнитным волнам.

В 1908 году Пуанкаре возвращается к вопросу о физических эффектах, которые возникают из-за существования гравитационных волн. В частности, он предсказывает [6], что из-за излучения гравитационных волн должно наблюдаться вековое уменьшение периода обращения в астрофизических бинарных системах. Именно это явление и обнаружили Джо Тэйлор и Рассел Халс в 1974 году при изучении двойного пульсара PSR B1913+16 [7]. Несмотря на то, что Пуанкаре правильно предсказал существование этих новых физических явлений, его рассуждения были скорее качественными – полноценной количественной теории у него не было.

Следующий важный шаг в создании релятивистской теории гравитации был сделан в 1912 году Максом Абрахамом (на фотографии). В своих работах Абрахам детально рассматривал вопрос о гравитационных волнах. Он доказал, что дипольного гравитационного излучения не существует – это следует из закона сохранения импульса и равенства гравитационных и инертных масс. Но довольно быстро время показало, что релятивистская теория гравитации Абрахама – это дорога в никуда. У Эйнштейна и Абрахама были на этот предмет жаркие споры, которые, возможно, помогли Эйнштейну осознать ограниченность всех релятивистских теорий гравитации как теории поля в плоском пространстве-времени Минковского.  Как писали в 1923 году Макс Борн и Макс фон Лауэ в некрологе Абрахаму, «он был достойным оппонентом, сражался достойным оружием и не старался замаскировать поражения причитаниями и не относящимися к делу аргументами… Он любил свой абсолютный эфир, свои уравнения поля, свой неподвижный электрон, как повзрослевший человек любит свою первую страсть, воспоминания о которой не затмит никакой последующий опыт».

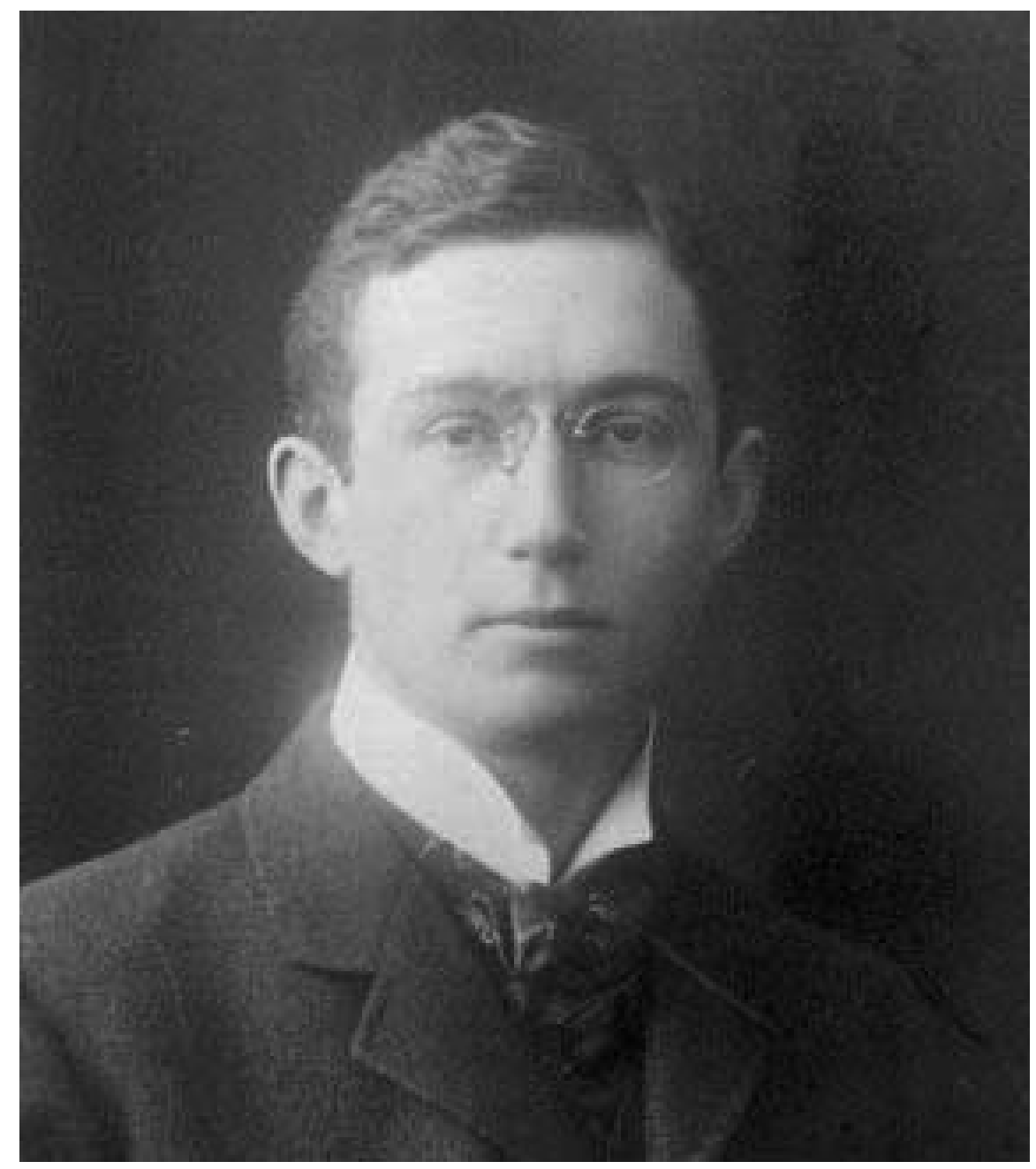

В 1915 году долгий путь Эйнштейна к правильной теории гравитации закончился успехом – созданием общей теории относительности. Вскоре, в феврале 1916 года, Эйнштейн написал Карлу Шварцшильду, что в новой теории нет никаких гравитационных волн, аналогичных электромагнитным волнам, и что это, возможно, связано с отсутствием гравитационного аналога электрического диполя (так как масса всегда положительна). Это удивительный вывод, так как для любого взаимодействующего поля ожидается существование волн. Мы не знаем, как Эйнштейн пришел к этому выводу, но можем только предположить следующее [8] (см. также книгу Daniel Kennefick , Traveling at the Speed of Thought:  Einstein and the Quest for Gravitational Waves).

Возможно, Эйнштейн хотел обобщить свои вычисления смещения перигелия Меркурия на более высокие порядки теории возмущения. Пост-ньютоновские поправки в общей теории относительности имеют дополнительную малость второго порядка по $\beta$ (отношение скорости планеты к скорости света). Эти поправки и обеспечивают правильное значение смещения перигелия Меркурия – первый триумф общей теории относительности. Если бы дипольное гравитационное излучение существовало, мы его увидели бы в вычислениях пост-ньютоновских поправок в третьем порядке по $\beta$ по сравнению с основным ньютоновским пределом. Эйнштейн это и сделал, наверное, и увидел, что никакого гравитационного излучения нет. Но, как уже было сказано, отсутствие дипольного гравитационного излучения было доказано Абрахамом из общих соображений. Как нам сейчас известно, в общей теории относительности, чтобы увидеть реакцию излучения, надо уйти на пять дополнительных порядков по $\beta$ от Ньютоновского предела.   Это трудная задача как технически, так и концептуально.

Вскоре Эйнштейн поменял свое мнение относительно гравитационных волн. Эйнштейн пытался найти приближенное решение своих уравнений, но большого прогресса не было в течение нескольких месяцев. Прогресс наступил, когда он по совету де Ситтера (на фотографии ниже Эйнштейн и де Ситтер обсуждают некоторую научную проблему в 1932 году) стал использовать так называемые гармонические координаты.

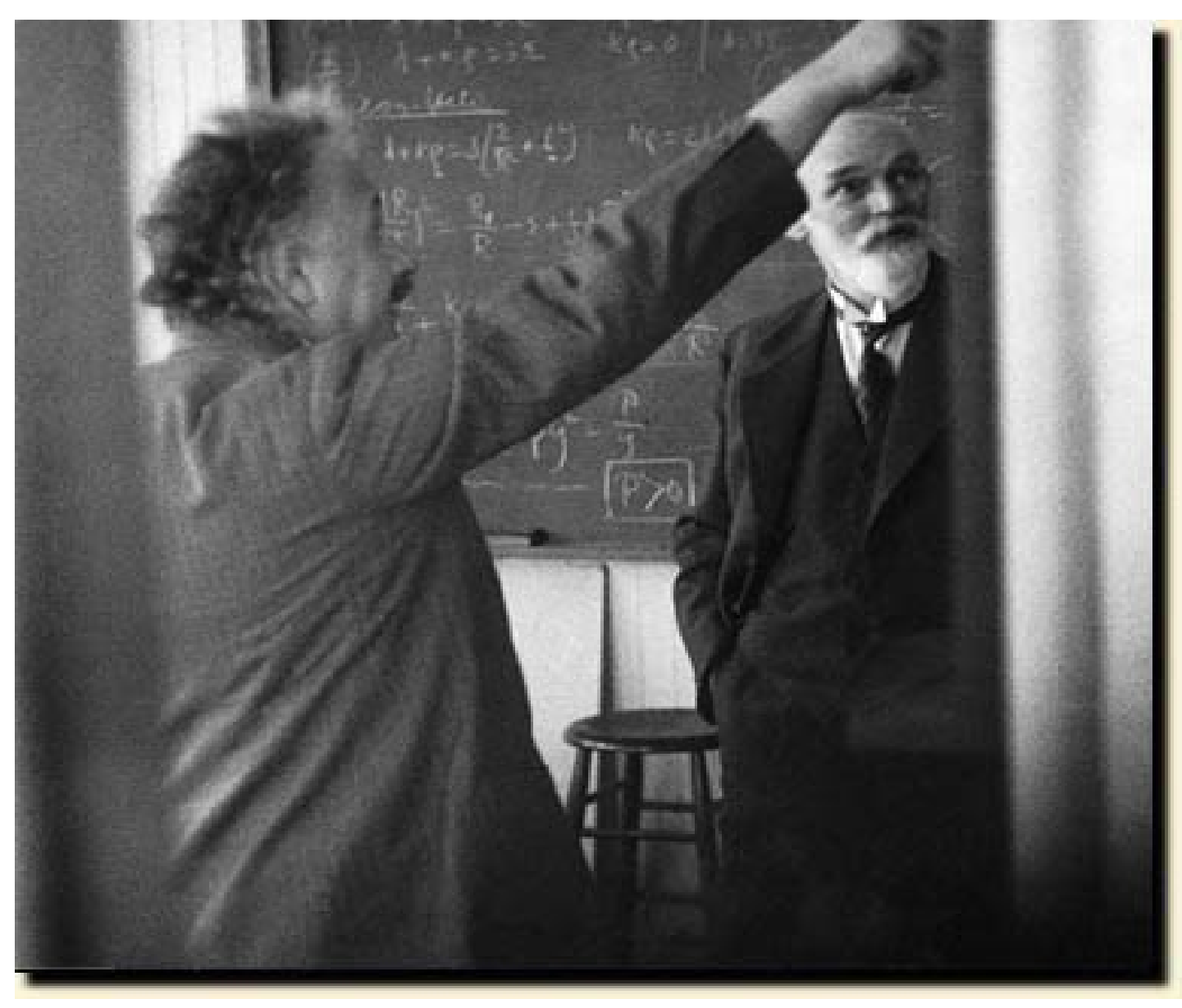

Эйнштейн в том же 1916 году получил свою знаменитую квадрупольную формулу для гравитационного излучения и установил существование трех видов гравитационных волн: продольно-продольных, продольно-поперечных и поперечно-поперечных (так потом их назовет немецкий математик Герман Вейль). При этом он показал (правда, используя неверные предположения), что первые два типа волн не переносят энергии и, таким образом, являются фиктивными, а не настоящими гравитационными волнами. Он дал правильную физическую интерпретацию этого факта. Дело в том, что гармонические координаты в некотором смысле сами соответствуют волнообразному движению. Найденные Эйнштейном фиктивные волны были всего лишь плоским пространство-временем Минковского, но воспринятым из координатной системы с волнообразным движением.

Вскоре после выхода работы Эйнштейна финский физик-теоретик Гуннар Нордстрём (на фотографии слева) использовал результаты Эйнштейна в своих вычислениях и получил странные результаты. Сначала Эйнштейн убеждал Нордстрёма в своей правоте, но Нордстрём упорствовал. Кроме того, вскоре Эйнштейн получил другие критические замечания от итальянского математика Туллио Леви-Чивиты (на фотографии справа).

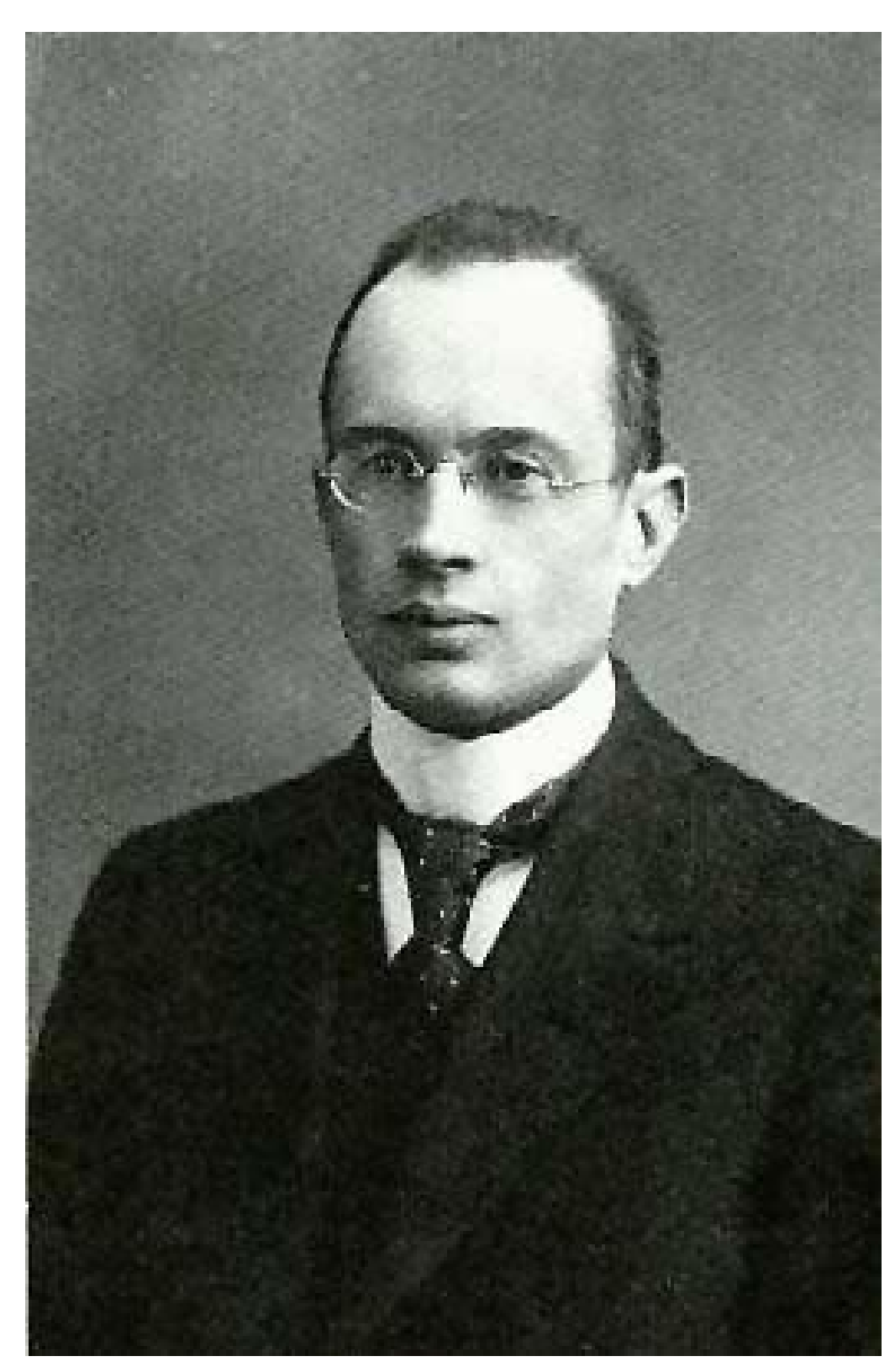
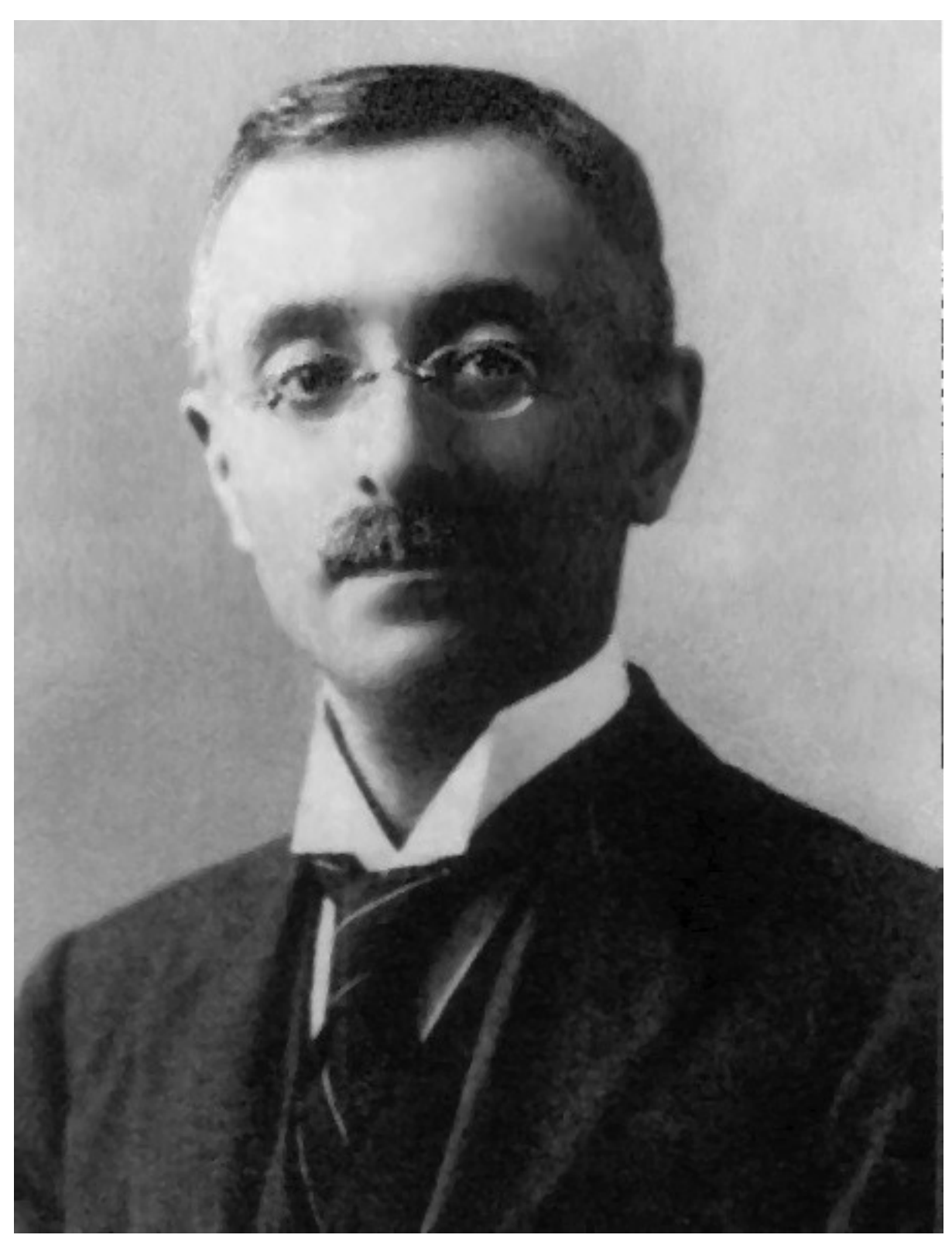

Леви-Чивита показал, что математическое выражение, которое Эйнштейн получил для описания энергии гравитационного поля, не являлось тензором (выражение не было ковариантным – вело себя при преобразованиях координат не так, как подобает истинной физической величине) и, таким образом, противоречило самому духу общей теории относительности, в которой физические величины являются тензорами (подчиняются определенным римановой геометрией законам преобразования при переходе от одних координат к другим). Псевдотензор энергии-импульса гравитационного поля Эйнштейна не имел правильных трансформационных свойств и, таким образом, зависел от выбора координат, а не только от гравитационного поля – весьма опасное свойство для физической величины. Поэтому Леви-Чивита сомневался, что псевдотензор Эйнштейна действительно имеет физический смысл.

Последующие события, казалось бы, подтвердили опасения Ливи-Чивиты. Сначала Шредингер показал, что, несмотря на наличие гравитационного поля вокруг сферически-симметричного источника, можно выбрать координаты так, что все компоненты псевдотензора Эйнштейна превратятся в нуль. Потом Бауэр показал обратное: в плоском пространстве-времени, т.е. в отсутствии гравитационного поля, можно так выбрать координаты, что компоненты псевдотензора Эйнштейна будут не нулевыми.

Эйнштейн вел интенсивную научную переписку со всеми участниками этих событий, а также с другими учеными того времени.  12 апреля 1918 года он пишет Гильберту: «Мой [комплекс энергии-импульса гравитационного поля] отвергается всеми как некошерный».

В 1918 году Эйнштейн пишет работу «О гравитационных волнах». Он исправляет некоторые ошибки (но не все) своей ранней работы 1916 года. В частности, коэффициент в квадрупольной формуле поменялся с $1/(24\pi)$ на $1/(80\pi)$. Но более существенно то, что Эйнштейн нашел ошибку в своих рассуждениях относительно псевдотензора гравитационного поля и получил правильное выражение для него. Эта ошибка являлась источником странных результатов, полученных  Нордстрёмом в своих вычислениях.

Что касается возражения Леви-Чивиты о нековариантном характере Эйнштейновского комплекса энергии-импульса гравитационного поля, Эйнштейн сослался на свой принцип эквивалентности для оправдания зависимости энергии гравитационного поля от системы отсчета. Принцип эквивалентности, который Эйнштейн называл «счастливейшей мыслью своей жизни», в первоначальной Эйнштейновской формулировке утверждает, что гравитационные и инерционные силы имеют одинаковую природу, подобно электрическим и магнитным силам, и ни один физический эксперимент не  сможет отличить гравитационное поле от ускорения системы отсчета локально. В лифте, который свободно падает, предметы ведут себя так же, как если бы не было однородного гравитационного поля и лифт был бы неподвижным.

Эйнштейн рассуждал так. Пусть в некоторой системе отсчета нет гравитационного поля. Следовательно, плотность энергии-импульса гравитационного поля в этой системе тождественно равна нулю. Если плотность энергии-импульса гравитационного поля является тензором, то во всех других системах координат он тоже будет равен нулю. Но если мы перейдем в равномерно-ускоренную систему отсчета, то появится сила инерции, которая, согласно принципу эквивалентности, не что иное, как однородное гравитационное поле, которое может совершать работу и, следовательно, обладает энергией. Поэтому, если мы хотим сохранить принцип эквивалентности, плотность энергии-импульса гравитационного поля должен зависеть от системы отсчета и не может быть тензором.

Большинство современных физиков отождествляют гравитацию с кривизной пространства-времени, математическим представлением которой является тензор Римана, и этим они отличаются от Эйнштейна. Например, в сентябре 1950 года в письме к Макс Фон Лауэ Эйнштейн прямо говорит, что, с эмпирической точки зрения, присутствие гравитационного поля означает отличие от нуля символов Кристоффеля (которые выражаются через метрический тензор и его производных первого порядка), а не отличие от нуля тензора Римана (фрагмент письма приведен в статье J.J. Stachel, How Einstein discovered general relativity: a historical tale with some contemporary morals). Однако, в 1967 году Питер Хавас показал [9], что Эйнштейновская трактовка принципа эквивалентности неприемлема: если взять координаты, в которых символы Кристоффеля равны нулю в некоторой точке пространства-времени, то простое введение криволинейных координат может привести к тому, что некоторые символы Кристоффеля станут отличными от нуля. Это, согласно Эйнштейновской интерпретации символов Кристоффеля, означало бы появление гравитационного поля даже несмотря на отсутствие ускорения - система отсчета как была локально-инерциальной, так таковой и осталась.

Общепризнано, что общая теория относительности – самая совершенная физическая теория. Забавно, что Эйнштейн создал эту замечательную теорию исходя из двух ложных предпосылок: принципа эквивалентности и желания обобщить принцип относительности на неинерциальные системы отсчета. Как писал Владимир Александрович Фок [10], «Констатировав факт равенства инертной и весомой массы, Эйнштейн переходит к рассмотрению поведения тел в «ускоренной системе отсчета» и приходит к выводу о равноправии всех систем отсчета – инерциальных и не инерциальных, что в свою очередь приводит его к требованию общей ковариантности уравнений. Опираясь на это требование, Эйнштейн приходит в конце концов к своим уравнениям тяготения. Первое и последнее звено этой цепи верны: несомненен факт равенства инертной и весомой массы, как несомненны и окончательные уравнения тяготения. Но промежуточные рассуждения Эйнштейна не выдерживают критики, так как они содержат ряд логических неувязок». В.А. Фок изображен на фотографии слева.

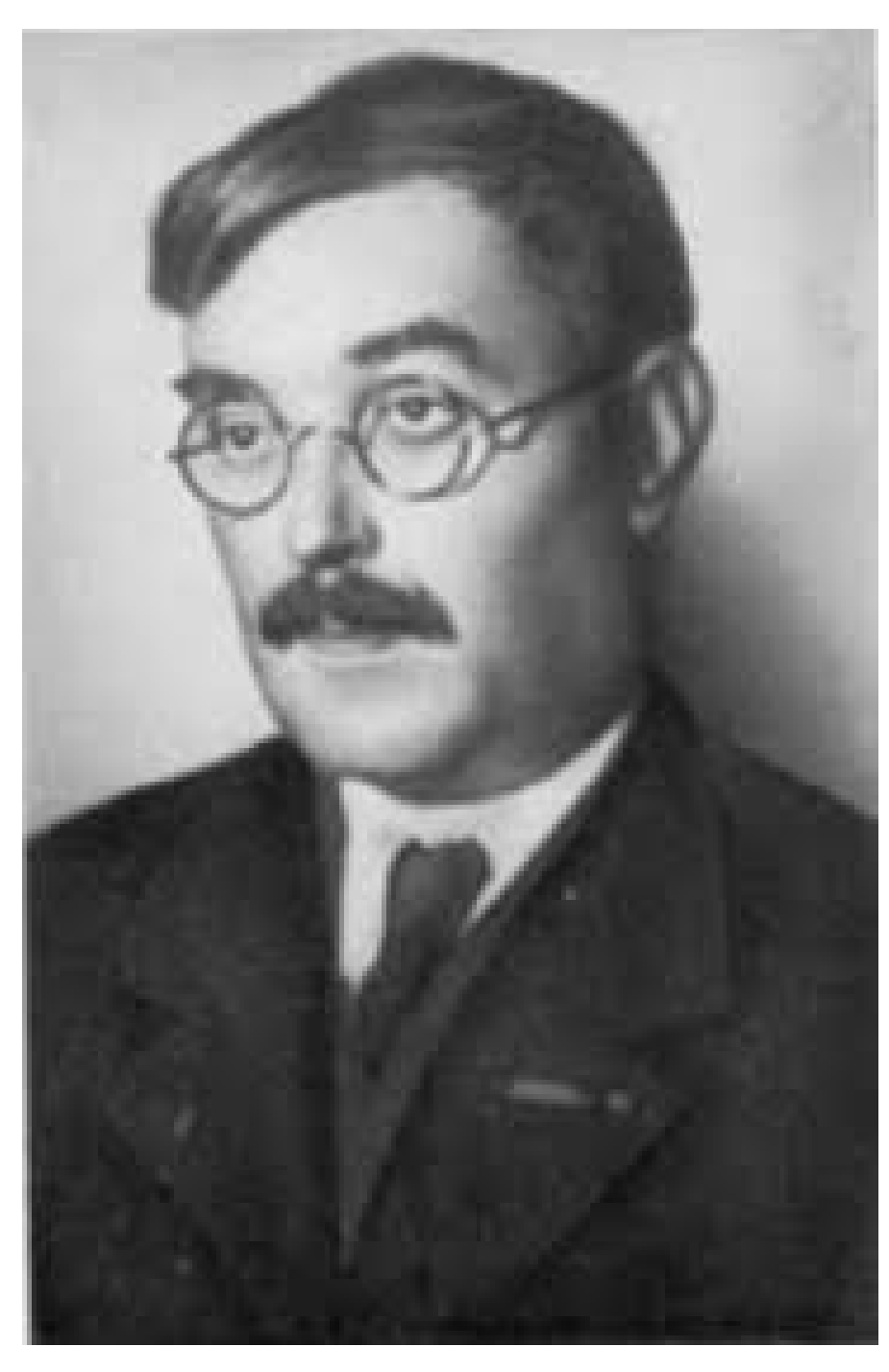

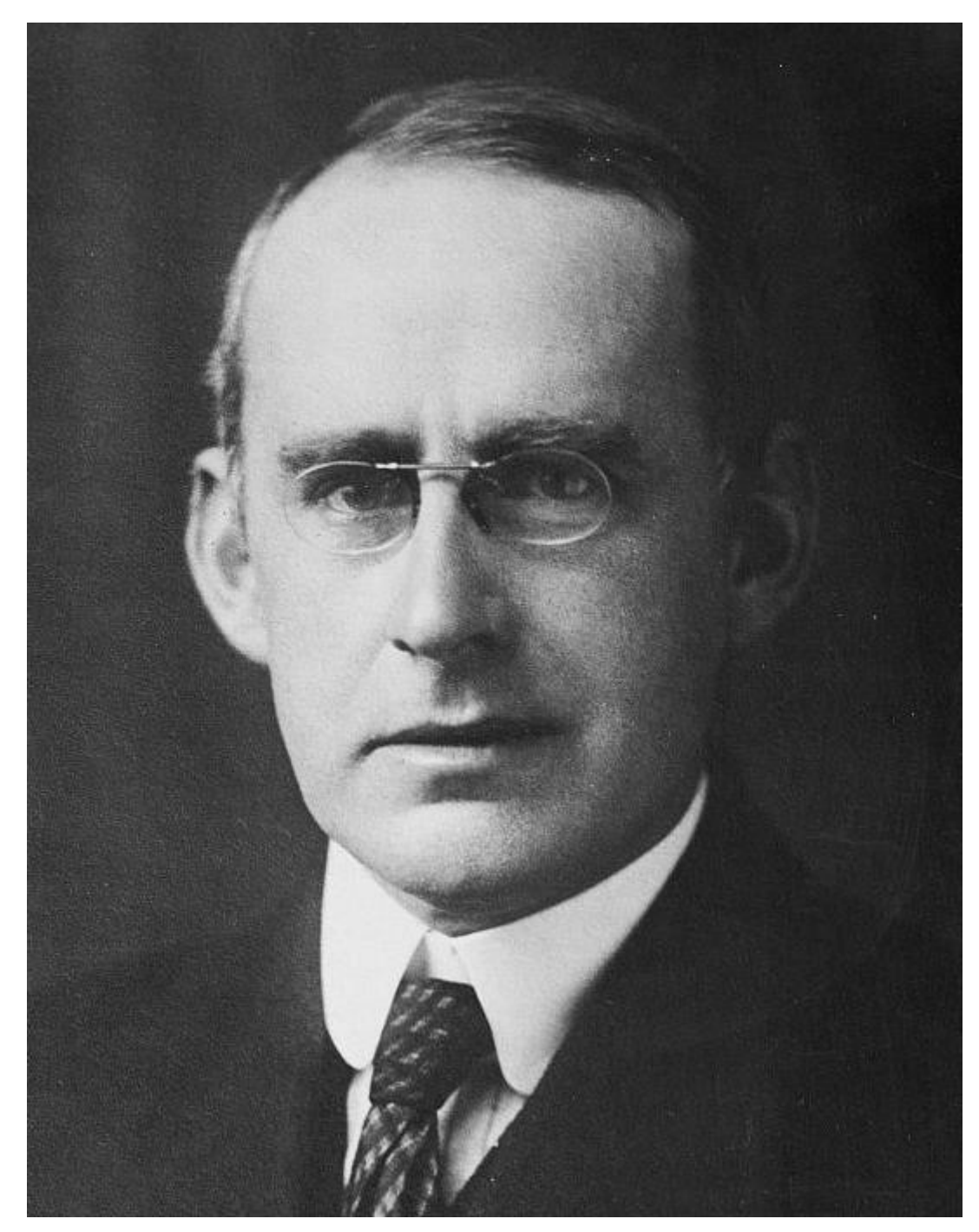

Но вернемся к гравитационным волнам. Следующий важный шаг как в развитии общей теории относительности, так и в теории гравитационных волн сделал английский астрофизик Артур Стэнли Эддингтон (на фотографии справа). Вместо того, чтобы работать с координатами, Эддингтон рассматривал волны кривизны. Заметим, что возможность существования волн кривизны предположил известный английский математик Уильям Кдиффорд еще в 1876 году [8]. Тем не менее, Клиффорд имел ввиду кривизну обычного трехмерного пространства, а не пространства-времени.

Подход Эддингтона подтвердил догадку Эйнштейна, что чисто координатные волны (на современном языке, «калибровочные степени свободы») могут распространяться с любой скоростью (по словам Эддингтона, «со скорость мысли»), и что они не переносят никакой энергии. Эддингтон в основном подтвердил и другие результаты Эйнштейна 1918 года, хотя он и нашел небольшую поправку: коэффициент в квадрупольной формуле вырос в два раза.

В последующие десятилетия вопрос о том, можно ли Эйнштейновскую квадрупольную формулу применить для реальных астрофизических систем, вызывал жаркие споры. Дело в том, что Эйнштейн получил свою формулу в линейном приближении. В этом приближении, и в системе координат, которую использовал Эйнштейн, линеаризованные уравнения гравитации принимают форму, которая аналогична уравнениям Максвелла для электромагнитного поля, и эта аналогия позволила Эйнштейну значительно упростить расчет. Но, поскольку общая теория относительности является нелинейной теорией, это линеаризованное приближение справедливо только для очень слабых полей, и в этом приближении двойная звездная система даже не образует связанного состояния. По этой причине Эддингтон считал, что линеаризованная теория недействительна для таких источников гравитационных волн, как бинарные звезды.

В 1941 году вышел второй том знаменитого курса Ландау и Лифшица (на фотографии). Они утверждали, что линеаризованный расчет, подобный Эйнштейновскому 1918 года, можно обобщить на случай двойной звездной системы, просто импортируя в расчет известные точные решения для движения звезд в отсутствие гравитационных волн. Таким способом они получили ту же самую квадрупольную формулу, что и Эйнштейн в 1918 году.

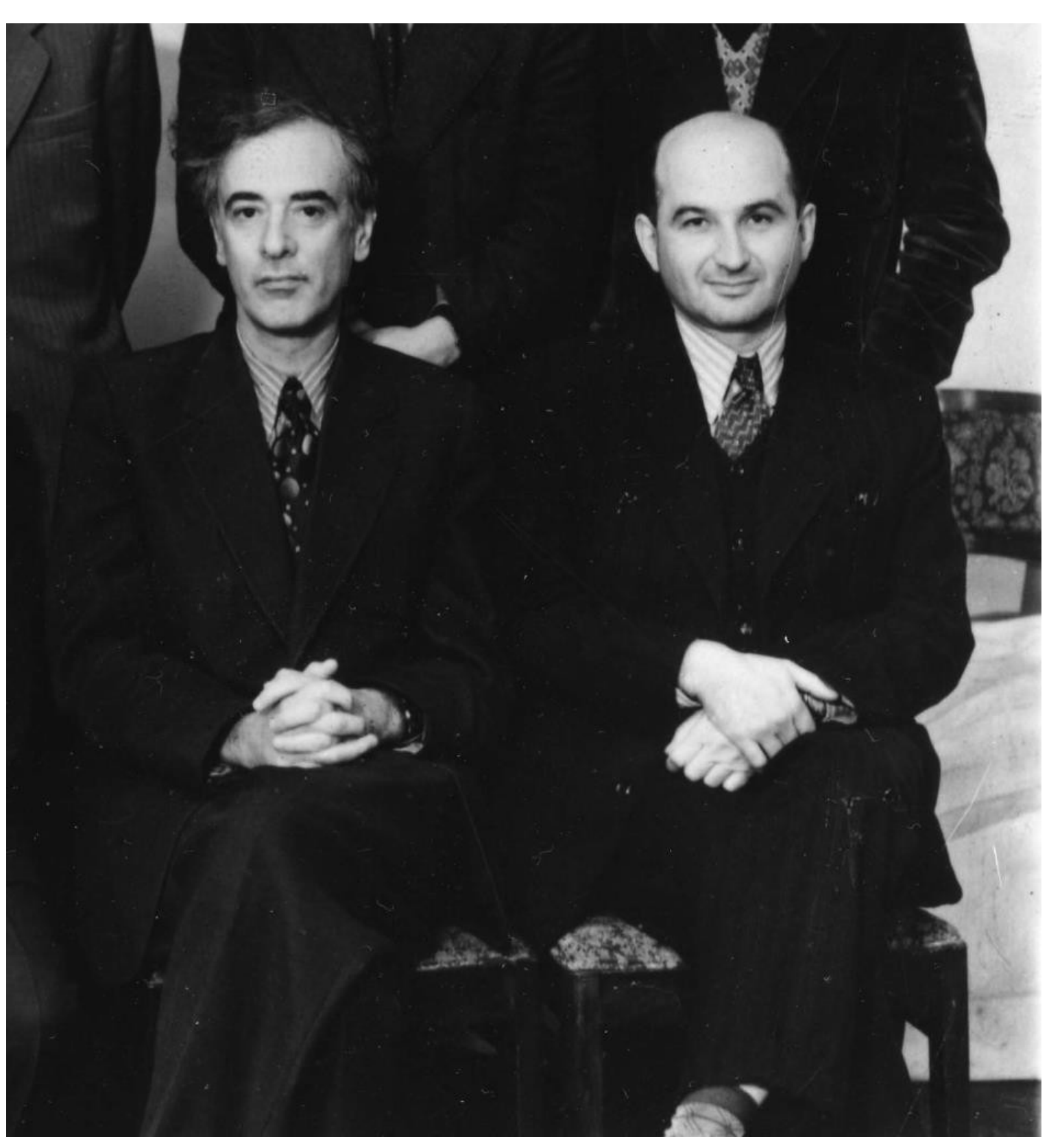

Аргумент Ландау и Лифшица об использовании правильной физическуй информации, даже если она математически несамосогласована, многие релятивисты нашли привлекательным с точки зрения физики, хотя скептицизм относительно справедливости квадрупольной формулы остался.

Понимая ограниченность линейного приближения, сам Эйнштейн, совместно со своим молодым сотрудником Натаном Розеном, в 1936 году принимается искать точное решение своих нелинейных уравнений в виде плоских гравитационных волн. Как бы они не старались, они не смогли это сделать без введения особенностей в компоненты метрики, описывающей гравитационную волну. В результате они пришли к выводу, что гравитационные волны вообще не существуют. Свою статью с названием «Существуют ли гравитационные волны?» они отправили в американский журнал Physical Review. Главный редактором этого журнала, который медленно, но верно становился одним из ведущих мировых физических журналов, тогда был малоизвестный американский физик Джон Торренс Тэйт. Эйнштейн и Розен до этого уже напечатали две статьи в этом журнале: одну про «кротовые норы», и вторую, совместно с Подольским, про так называемый парадокс Эйнштейна-Подольского-Розена в квантовой механике. Обе статьи были напечатаны без промедления и без всякого рецензирования, и время подтвердило адекватность такого решения редактора – сейчас обе статьи хорошо известны, особенно вторая.

В случае со статьей о гравитационных волнах Тэйт почувствовал неладное, и больше месяца она недвижимо лежала у него на столе [11]. Наконец, он послал статью рецензенту. Рецензентом был, как сейчас выясняется, Говард Перси Робертсон [12], выдающийся американский математик и физик. Робертсон был одним из крупнейших в мире специалистов в области общей теории относительности и космологии. Он понял физический смысл сингулярностей в решении Эйнштейна-Розена: при особом выборе системы координат эти сингулярности располагались на оси цилиндра и, таким образом, соответствовали источнику гравитационных волн. Вместо того чтобы доказать, что гравитационные волны не существуют, Эйнштейн и Розен открыли цилиндрические гравитационные волны, как точные решения уравнений Эйнштейна.

Через некоторое время изумленный Эйнштейн получает письмо Тейта [11]: «Уважаемый профессор Эйнштейн. Возвращаю Вам статью о гравитационных волнах, написанную Вами совместно с д-ром Розеном, с некоторыми комментариями рецензента. Прежде чем публиковать эту работу, я был бы рад услышать Ваше мнение по поводу различных комментариев и критических замечаний рецензента. Искренне Ваш, Джон Т. Тэйт, редактор».

Разгневанный Эйнштейн отвечает молниеносно: «Мы, г-н Розен и я, направили рукопись в вашу редакцию для публикации и не давали разрешения на ознакомление с ней специалистов до ее выхода в свет. Я не вижу причин, по которым я должен реагировать на комментарии вашего анонимного эксперта, тем более что они явно ошибочны. Посему я намерен опубликовать эту работу в другом журнале. С уважением, А. Эйнштейн».

Несмотря на огромную славу и популярность Эйнштейна, Тэйт был непреклонен: «Уважаемый д-р Эйнштейн. Весьма сожалею, что обстоятельства привели Вас к решению опубликовать статью, написанную совместно с д-ром Розеном, в другом журнале. Возможно, я лично виноват в этом, потому как полагал, что Вы знакомы с правилами публикаций Американского физического общества, и что Вы воспримете замечания нашей редколлегии как должное и отнесетесь к ним соответствующим образом. Я не могу принять к публикации в Physical Review работу, авторы которой не согласны на рассмотрение ее редколлегией перед публикацией. Я полагал, что Вам это известно, иначе я сразу же вернул бы статью без рассмотрения. Сожалею, что Вы сочли замечания редколлегии ошибочными и не заслуживающими реакции. Искренне Ваш, Джон Т. Тэйт, редактор».

На фотографиях ниже показаны Робертсон (слева), Тэйт (справа) и Эйнштейн (в середине).

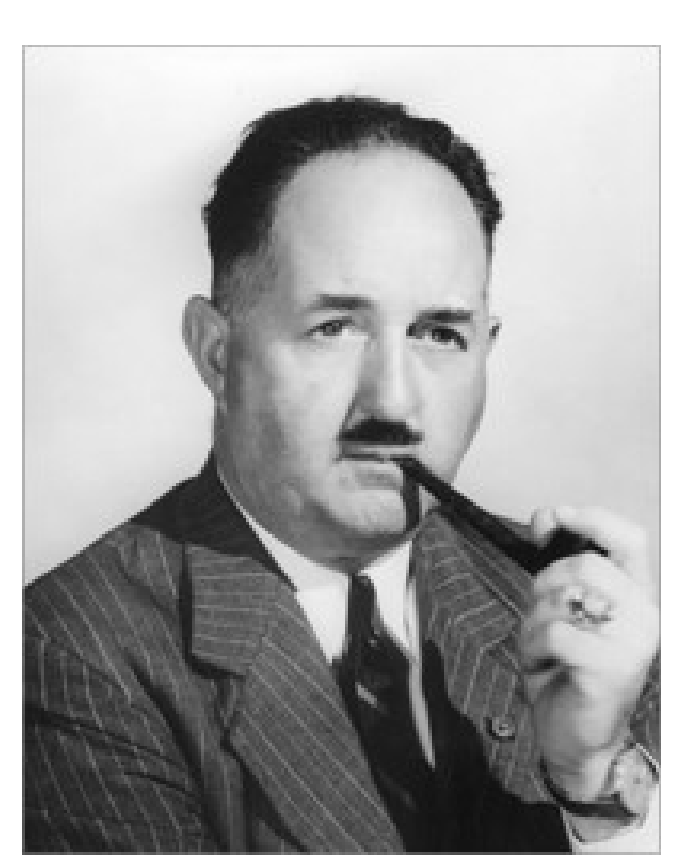

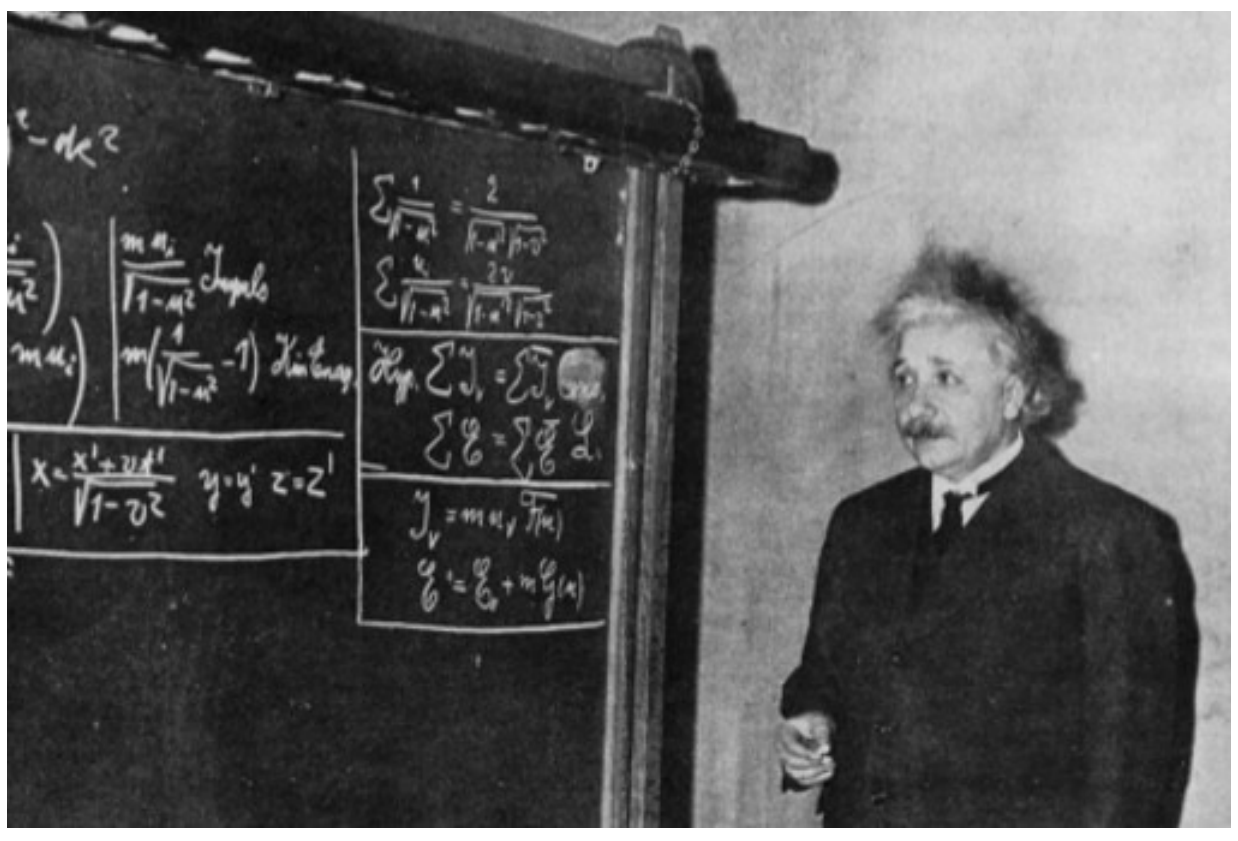

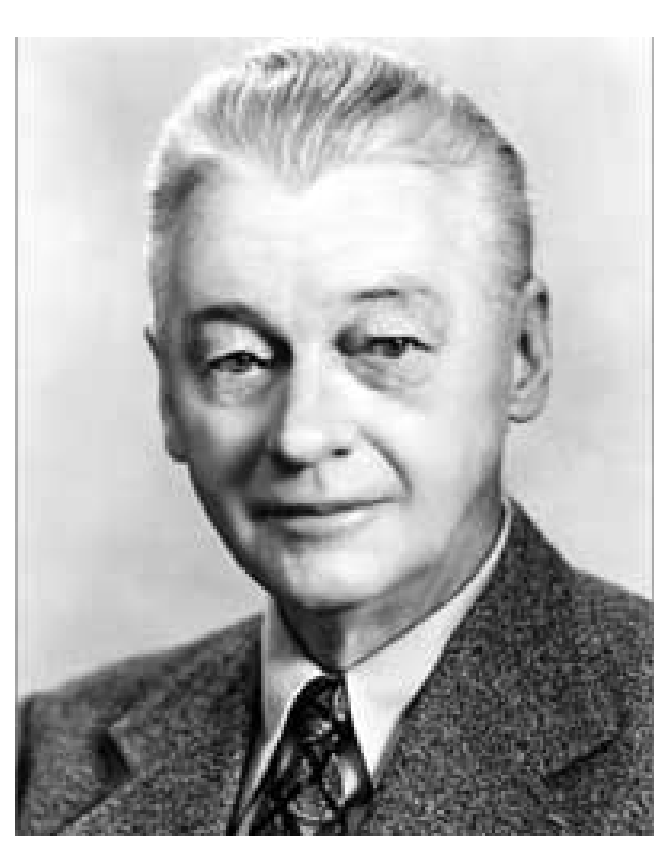

Эйнштейну все-таки удалось избежать публикации ошибочной работы. Разумеется, он не стал разбираться с замечаниями рецензента, а сразу послал работу в другой журнал (Journal of the Franklin Institute). Там статью сражу же приняли, но до ее фактической публикации произошли следующие события [12].

К Эйнштейну приехал другой ассистент – Леопольд Инфельд (на фотографии он вместе с Эйнштейном), заменив Натана Розена, уехавшего работать в Советский Союз. Поначалу Эйнштейну удалось убедить своего нового помощника, что его с Розеном аргументы верны, и, следовательно, гравитационные волны не существуют. Инфельд даже придумал свою версию доказательства того, что гравитационные волны не существуют. Но к этому времени в Принстон из отпуска вернулся Робертсон и, подружившись с Инфельдом, поделился с ним своим скептицизмом в отношении результата Эйнштейна. Они вместе шаг за шагом просмотрели Инфельдовскую версию доказательства и обнаружили ошибку. Инфельд рассказал об этом Эйнштейну. Эйнштейн к удивлению ответил, что сам случайно нашел ошибку в своих рассуждениях. Статья Эйнштейна и Розена была спешно переделана и вышла под названием «О гравитационных волнах».

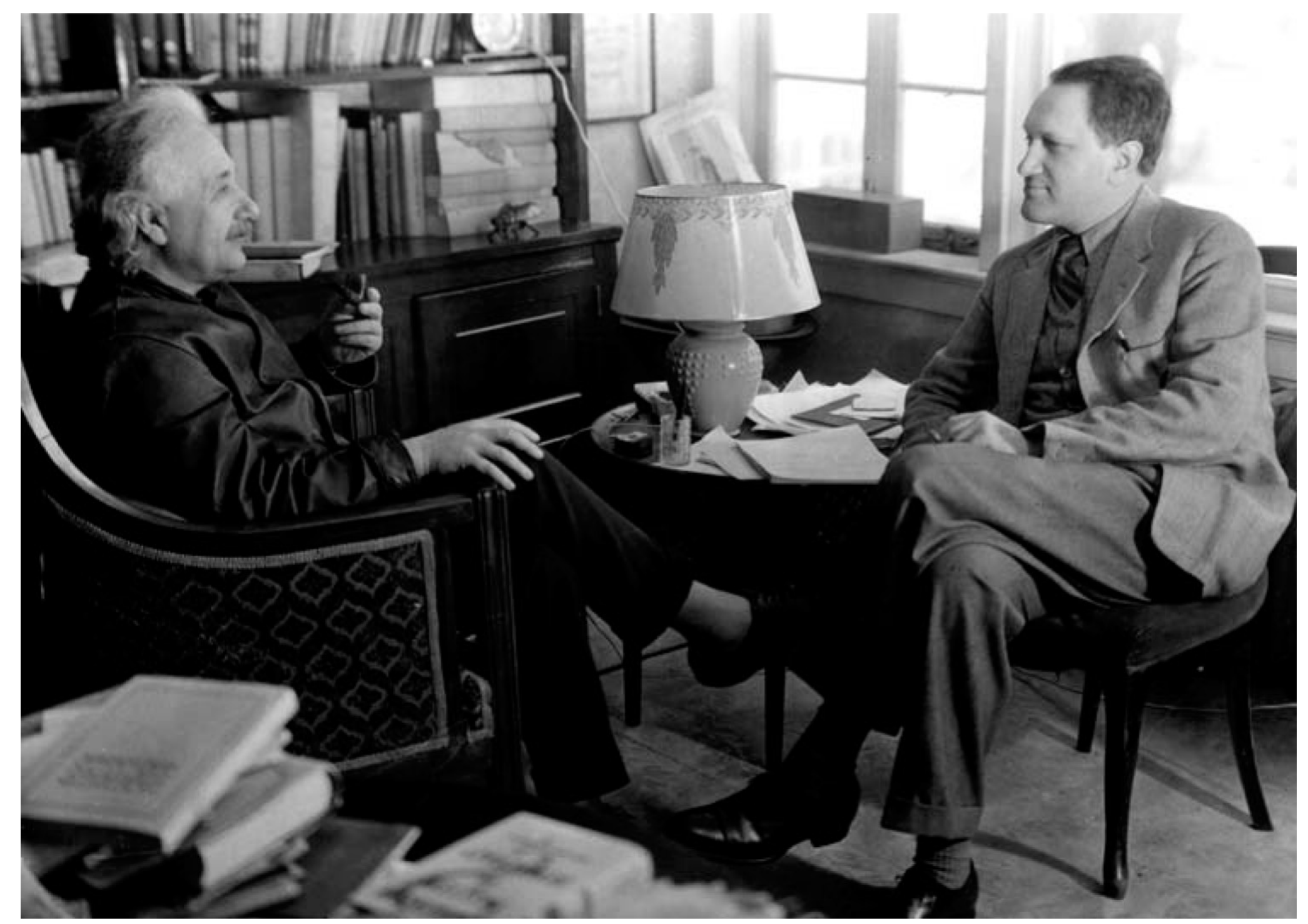

На первый взгляд, заявление Эйнштейна, что он сам нашел ошибку как раз накануне разговора с Инфельдом, не вызывает доверия. Но, наверное, все так и было. Сравнительно недавно была обнаружена рукопись неопубликованной работы Эйнштейна (см. книгу Daniel Kennefick, Traveling at the Speed of Thought: Einstein and the Quest for Gravitational Waves), которую он начал сразу после окончания его работы с Розеном. Эйнштейн искал другие примеры того, что решениям линеаризованных уравнений гравитации не всегда отвечают точные решения. Рукопись неожиданно обрывается на 11-ой странице при доказательстве отсутствия вращательно-симметричного стационарного решения точных уравнений Эйнштейна. Видимо, именно в этот момент Эйнштейн и понял ошибочность своих аргументов.

Косвенно это подтверждается и другим фактом. Согласно Инфельду (см. упомянутую книгу Дэниела Кеннефика), Эйнштейн должен был читать лекцию в Принстоне о невозможности гравитационных волн всего лишь через день после своего открытия неверности своего доказательства. Он еще не разговаривал с Робертсоном и не знал, что на самом деле он открыл цилиндрические гравитационные волны. Поэтому он был вынужден читать лекцию о недействительности его собственного доказательства, заключая: «Если вы спросите меня, существуют ли гравитационные волны или нет, я должен ответить на это, что я не знаю. Но это очень интересная проблема».

Хотя точное решение уравнений гравитации, соответствующее цилиндрическим гравитационным волнам, сейчас носит название «Волна Эйнштейна-Розена», на самом деле это решение открыл в 1925 году - за 12 лет до Эйнштейна и Розена - австрийский физик Гвидо Бэк (на фотографии слева). Но работа Бэка прошла незамеченной. О ней помнил его студент Питер Хавас (на фотографии справа), но он стал заметен в этой области только к концу пятидесятых годов прошлого века.

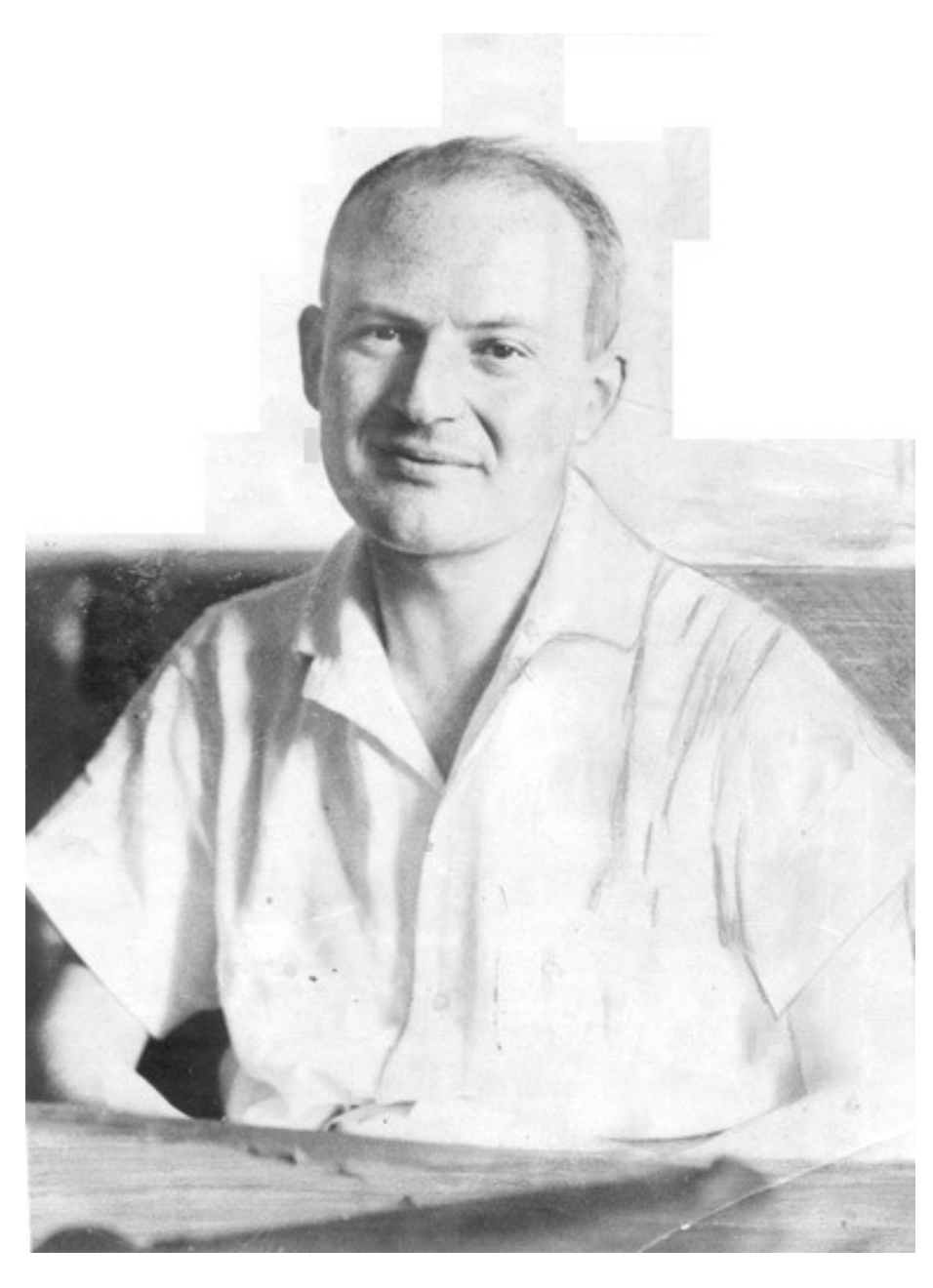

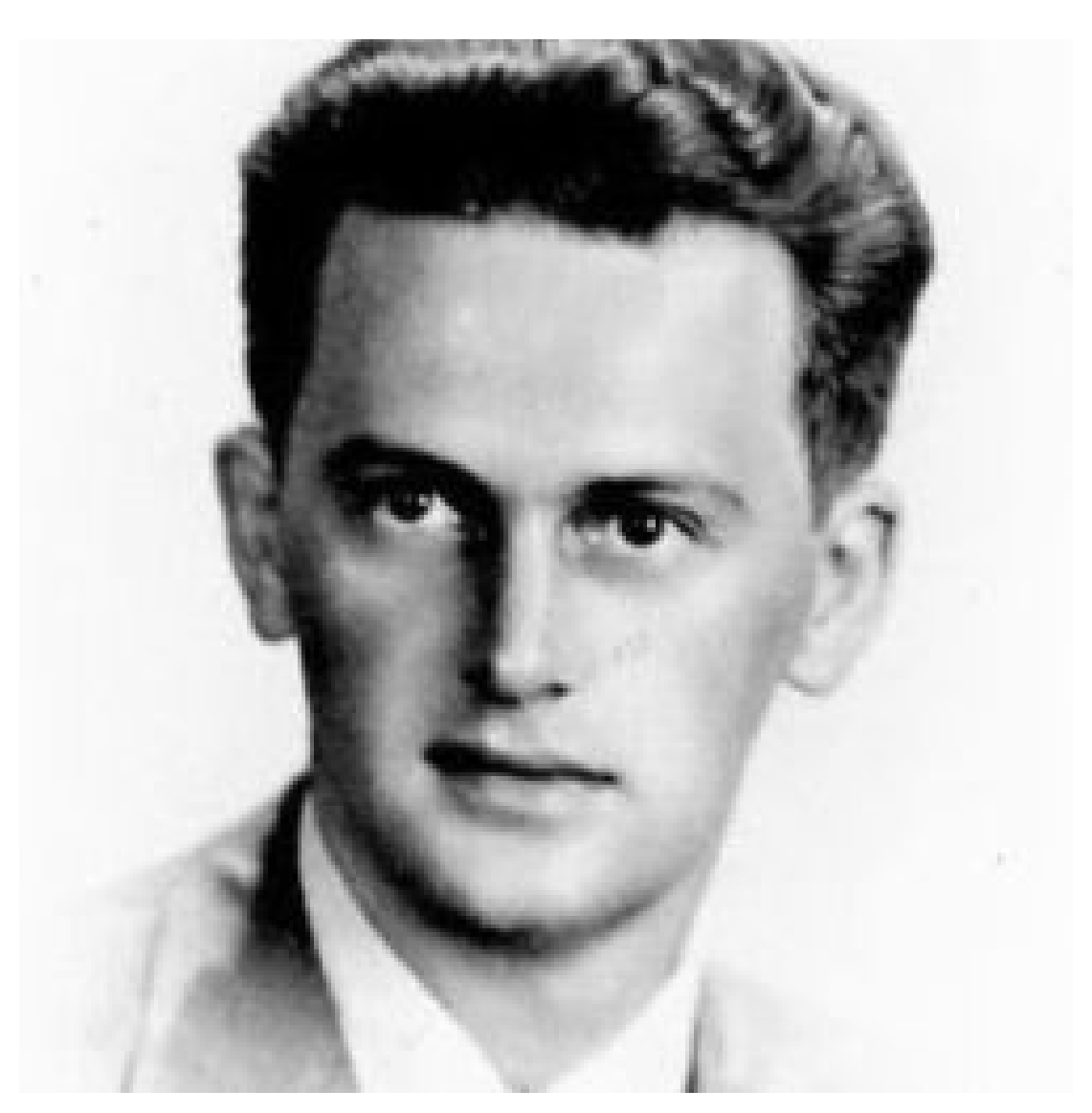

Интересно, что Гвидо Бэк в 1934 году приехал в Советский Союз и до 1937 года работал в Одесском университете.  Он многое сделал для становления теоретической физики в Одесском университете. Как он сам вспоминал, он был полностью измотан работой, так как работал по 18 часов в сутки. Он читал все лекции по курсам теоретической физики и руководил научной работой около пяти студентов. Студенты настолько доверяли Бэку, что начали жаловаться ему на философию и диалектический материализм, которые им преподавали в университете. Даже Ландау как-то признался Бэку: «Да, сейчас их учат военному делу в университете. Это хорошая вещь. Это может помочь защитить страну, если на нее нападут. А какая польза от диалектического материализма, я не знаю». В те времена эти разговоры были смертельно опасными, и, если бы Бэк, к огорчению своих студентов, не уехал в 1937 году из страны, вряд ли он избежал бы репрессий.

Интересно, что даже в опубликованной версии своей работы с Розеном Эйнштейн все же включил обсуждение возможности того, что бинарные звезды не будут излучать гравитационные волны несмотря на то, что его же квадрупольная формула указывает, что они будут. Леопольд Инфельд впоследствии всегда настаивал на том [13], что это и есть последнее слово Эйнштейна по данному вопросу, и по-видимому, если учитывать опубликованные работы Эйнштейна, так и есть.

Не только Эйнштейн сомневался в существовании гравитационных волн как реального физического явления. Розен и Инфельд выдвинули ряд аргументов, что орбитальная энергия двойной звезды не будет уменьшатся в результате испускания гравитационных волн. На этот счет Герман Бонди также высказал серьезные сомнения [13], утверждая, что аналогия с электромагнетизмом на самом деле указывает именно на это. В электродинамике ускоренные заряды излучают электромагнитное излучение, и нечто подобное ожидается в случае гравитации. Но какие массы считать ускоренными в общей теории относительности? По мнению Бонди, инерциальная частица в общей теории относительности — это частица, движущаяся по геодезической. А ускоренная частица — это та, которая не движется по геодезической из-за действия на нее не гравитационной силы. Поскольку звезды в гравитационных бинарных системах движутся вокруг друг друга по геодезическим локального пространства-времени, они в смысле Бонди не ускорялись и, следовательно, не могли излучать гравитационные волны.

Сегодня гравитационная физика находится в авангарде научного прогресса. Большие деньги тратятся на гравитационно-волновые эксперименты. Святой Грааль теоретической физики состоит в том, чтобы объединить ядерные и электромагнитные силы с гравитацией. Курс общей теории относительности является обязательным атрибутом образования студентов-физиков. Но это не всегда было так. Несмотря на свою существенную математическую полноту уже в 1916 году, до примерно шестидесятых годов прошлого века общая теория относительности развивалась очень медленно и изолированно от всей остальной физики.

Та пропасть, которая в то время разделяла релятивистов (исследователей в общей теории относительности) от остального физического сообщества наиболее отчетливо видна из письма Ричарда Фейнмана своей жене, написанного с конференции по общей теории относительности и гравитации в Варшаве в 1962 году (письмо можно найти в книге Р.Ф. Фейнман, Ф.Б. Мориниго, У.Г. Вагнер, Фейнмановские лекции по гравитации): «Я ничего не получил на этой конференции. Я не узнал ничего нового. Поскольку в этой области нет экспериментов, эта область науки находится в неактивном состоянии, так что только очень немногие из лучших людей работают в ней. Результат состоит в том, что здесь имеется огромное количество дурмана, и это сказывается неблагоприятным образом на моем артериальном давлении: такие бессмысленные вещи говорятся и серьезным образом обсуждаются, что я спорю с участниками вне формальных сессий, скажем, на ланче, всякий раз, когда кто-либо задает мне вопрос или начинает мне рассказывать о своей "работе". Эта "работа" всегда является: А) полностью непонятной, Б) неясной и неопределенной, В) в чем-то частично правильной, но то, что ясно и самоочевидно, тем не менее получается с помощью длинного и трудного анализа и представляется как важное открытие, или это Г) некоторое заявление, основанное на глупости автора, относительно очевидного и правильного факта, принятого и проверенного много лет назад; фактически, заявление, которое является неверным (что хуже всего, так как никакие доводы не убедят глупца), Д) попытка сделать что-либо, вероятно, невозможное, но определенно не несущей никакой пользы, и эта попытка, как в конце концов обнаруживается, приводит к провалу или, очевидным образом, является неверной. В эти дни проводится огромная "деятельность в этой области", но эта деятельность главным образом состоит в демонстрации того, что предыдущая "деятельность" кого-то еще приводит к ошибке, или не приводит ни к чему полезному, или приводит к чему-то, что подает надежды. Это напоминает группу червяков, пытающихся вылезти из бутылки, переползающих один через другого. И не потому, что задача трудна, а потому, что лучшие люди занимаются другими вещами. Напомни мне не ездить больше ни на какие конференции по гравитации!»

Фейнман конечно преувеличивает. Тем не менее, пройдет еще несколько десятилетий [14], прежде чем основная теоретическая физика полностью откажется от предрассудков против сообщества релятивистов. Но перелом наступил уже в 1957 году на конференции в Чапел Хилл, США, на которой Фейнман тоже присутствовал. Что касается гравитационных волн, этот перелом связан с именами Феликса Пирани (на фотографии слева), того же Фейнмана (на фотографии в середине) и Германа Бонди (на фотографии справа).

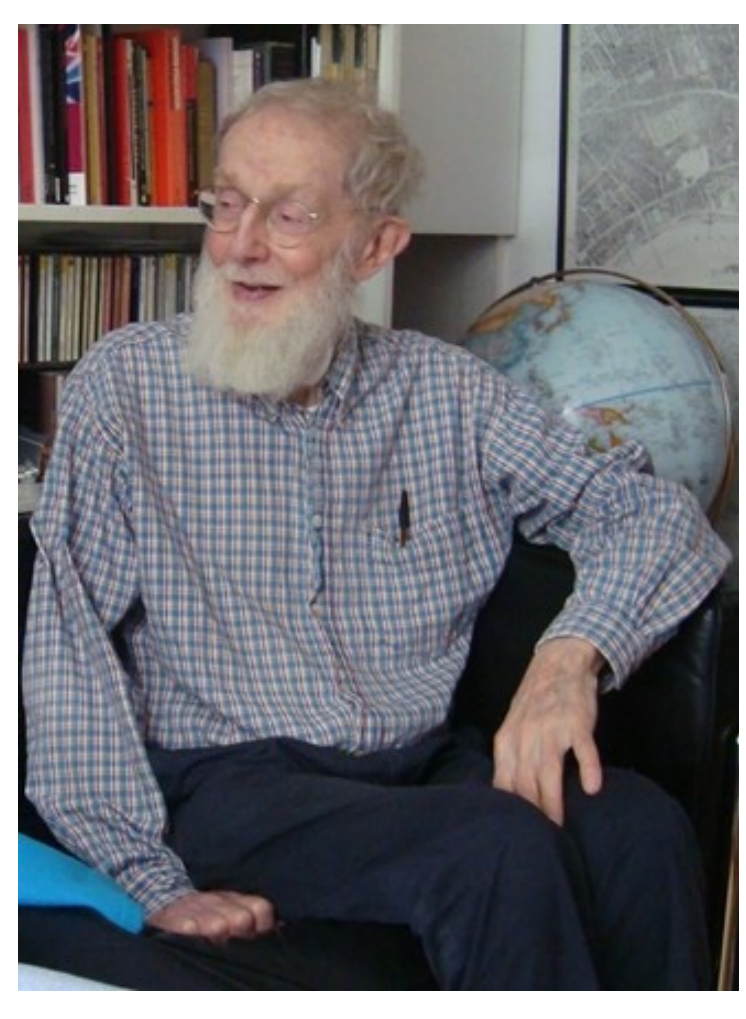
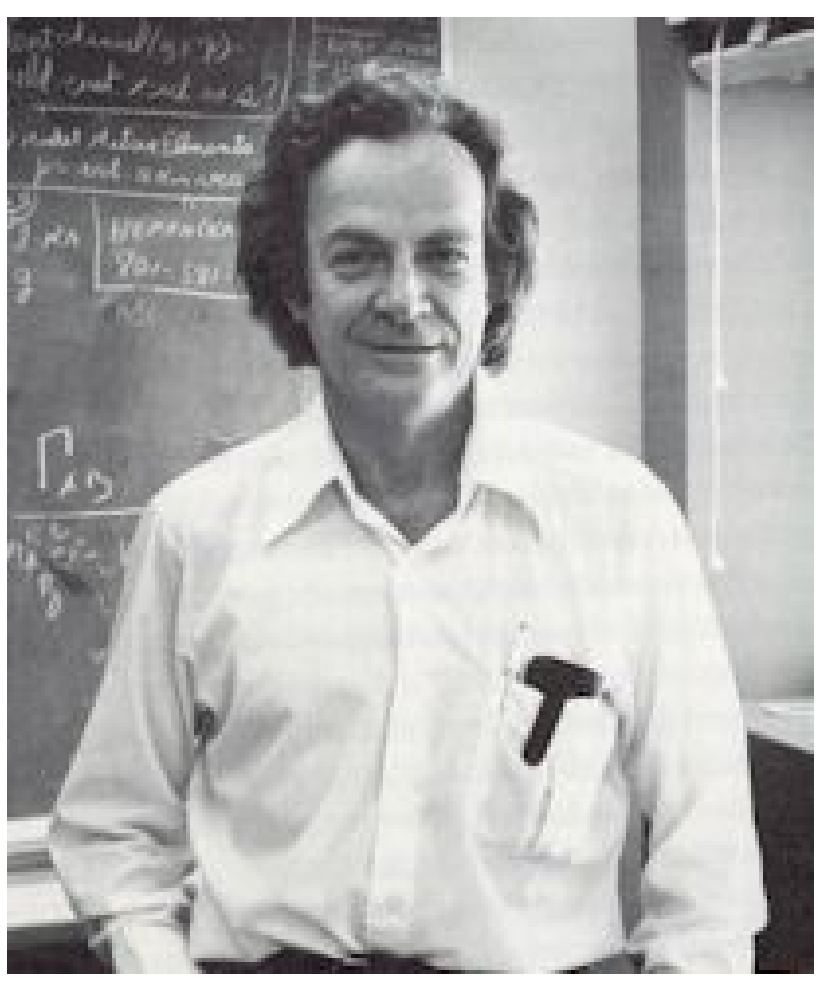
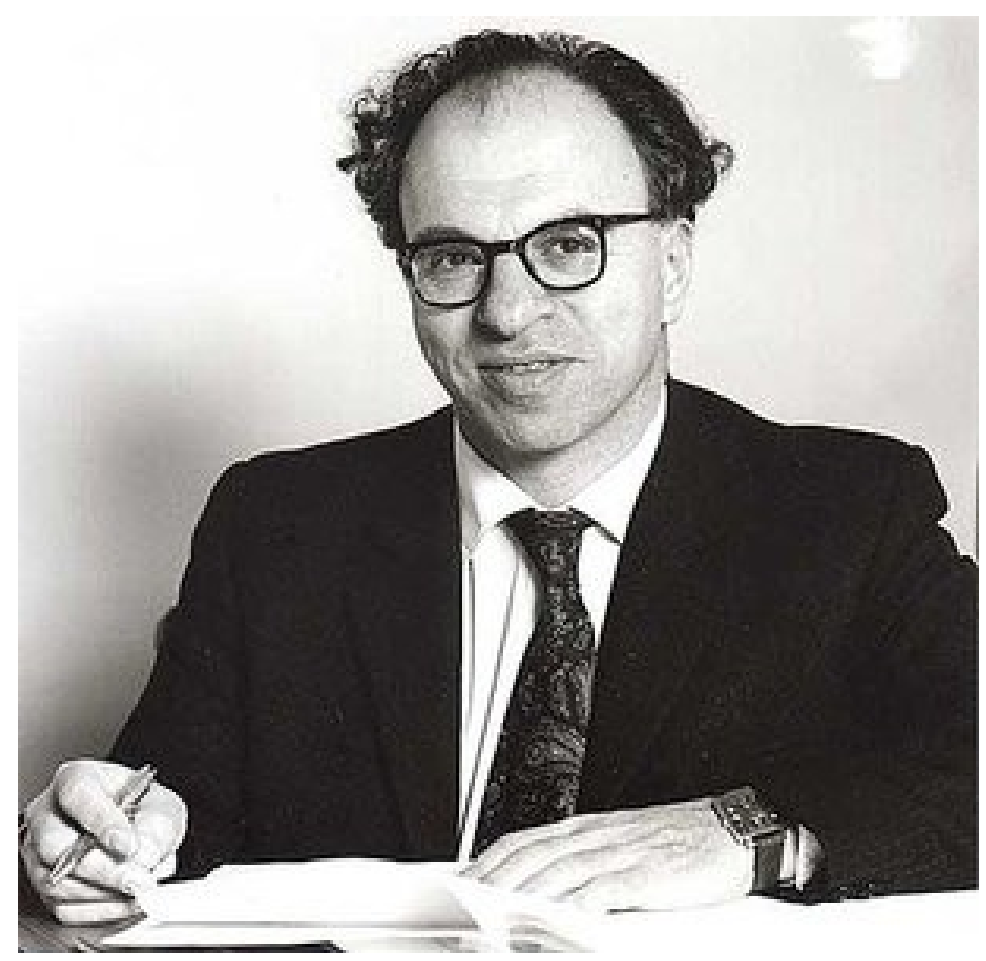

Сомнения релятивистов того времени в реальности гравитационных волн восходят по своим корням к замечанию Леви-Чивиты, что плотность энергии-импульса гравитационного поля не является тензором. На самом деле существует простой аргумент, предложенный Робертом Герохом [15], что нет такого тензорного поля, которое можно было бы интерпретировать как тензор энергии-импульса гравитационного поля.

Представим, что такой тензор есть. Тогда он либо добавляется к тензору энергии-импульса материи в правой части уравнении Эйнштейна, либо нет. В первом случае такой тензор должен зануляться для так называемых Риччи-плоских метрик, так как для таких метрик левая часть уравнении Эйнштейна зануляется. Но пространство-время, соответствующее чисто гравитационным степеням свободы, в том числе гравитационным волнам, как раз Риччи-плоское. Таким образом, такой тензор не может описывать энергию-импульс гравитационных степеней свободы. Неудивительно, что Розен и другие релятивисты часто приходили к заключению, что гравитационные волны не обладали энергией.

Во втором случае, когда предполагаемый тензор не добавляется к тензору энергии-импульса материи в правой части уравнении Эйнштейна, ситуация еще хуже. Из уравнения Эйнштейна, через тождество Бьянки, следует, что дивергенция тензора энергии-импульса материи равна нулю, то есть энергия-импульс материи сохраняется сам по себе, и не может быть никакого обмена энергией между материей и гравитационным полем.

Тот факт, что гравитационная энергия порождает концептуальные проблемы в общей теории относительности, является следствием того, что в этой теории гравитация «геометризируется» – в общей теории относительности гравитация понимается как инерционный эффект в искривленном пространстве-времени, а не как сила, совершающая работу, которую можно интерпретировать как изменение гравитационной энергии.

Малоизвестным фактом является то, что Ньютоновскую гравитацию тоже можно геометризировать, и после этого в соответствующей теории, называемой теорией Ньютона-Картана, возникают точно такие же проблемы [15] с определением гравитационной энергии, как в общей теории относительности. Чтобы в таких геометризованных теориях ввести понятие плотности гравитационной энергии, мы должны частично «дегеометризировать» гравитацию, то есть ввести дополнительную структуру [15] - некую привилегированную систему отсчета для определения инерционного движения. Тогда гравитационные эффекты появляются как отклонение от определенного таким образом инерциального движения. Эйнштейн в споре с Леви-Чивитой все-таки оказался прав – чтобы говорить осмысленно о плотности гравитационной энергии, нужны именно псевдотензоры, а не тензоры.

Но в том далеком 1957 году все это еще не было ясно, и релятивисты яростно и бесплодно продолжали спорить о реальности гравитационных волн. После конференции в Чапел Хилле ситуация радикально изменилась. Заметным событием этой конференции стал так называемый «аргумент скользящей бусинки». По-видимому, его независимо предложили [8] (см. также книгу П. Феррейра, Идеальная теория: битва за общую теорию относительности) Герман Бонди и Ричард Фейнман, который прибыл на конференции на день позже, пропустив презентацию Бонди. Особенно ясно этот аргумент изложил Фейнман. Идея состоит в том, что, если взять стержень, на который на некотором расстоянии надеты две бусинки, то проходящая гравитационная волна на стержень практически не повлияет, так как стержень жесткий, а бусинки начнут смещаться вдоль стержня подобно прыгающим вверх и вниз буйкам на поверхности моря. Если бусинки движутся вдоль стержня с некоторым трением, будет выделяться тепло, и стержень будет греться. Энергии для нагрева стержня неоткуда взяться, кроме как от гравитационной волны. Многие восприняли это как убедительный аргумент, что гравитационные волны могут переносить энергию, но не все, как убедился разочарованный Фейнман на конференции в Варшаве через несколько лет.

Хотя Бонди и/или Фейнман обычно получают заслуженное признание за свой вклад, на самом деле концептуальный сдвиг в решении проблемы гравитационных волн, включая вопросы их реальности, способов генерации и регистрации, связан с именем Феликса Пирани [16]. Именно после доклада Пирани, в котором были изложены его результаты о влиянии гравитационной волны на движение материальных тел, Бонди сменил свою скептическую позицию относительно реальности гравитационных волн. Фейнман тоже опирался в своих рассуждениях на результаты Пирани.

На конференции в  Чапел Хилле присутствовал один человек, который, под влиянием  аргумента скользящей бусинки, решил, что настало время перевести проблему гравитационных волн в практическую плоскость и построить детектор для их регистрации. Звали этого человека Джозеф Вебер (на фотографии слева он показан вместе со своим детектором гравитационных волн).

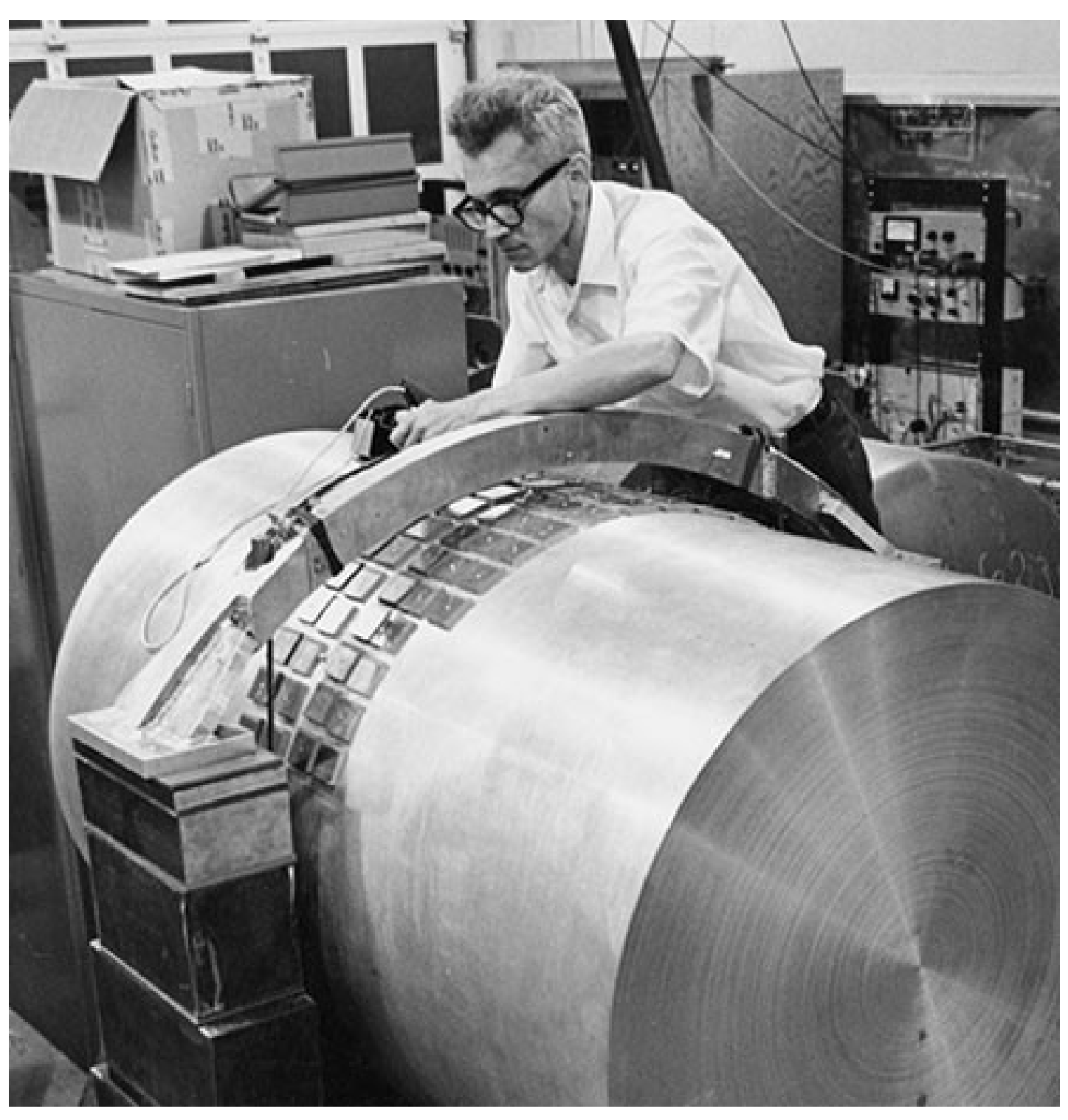

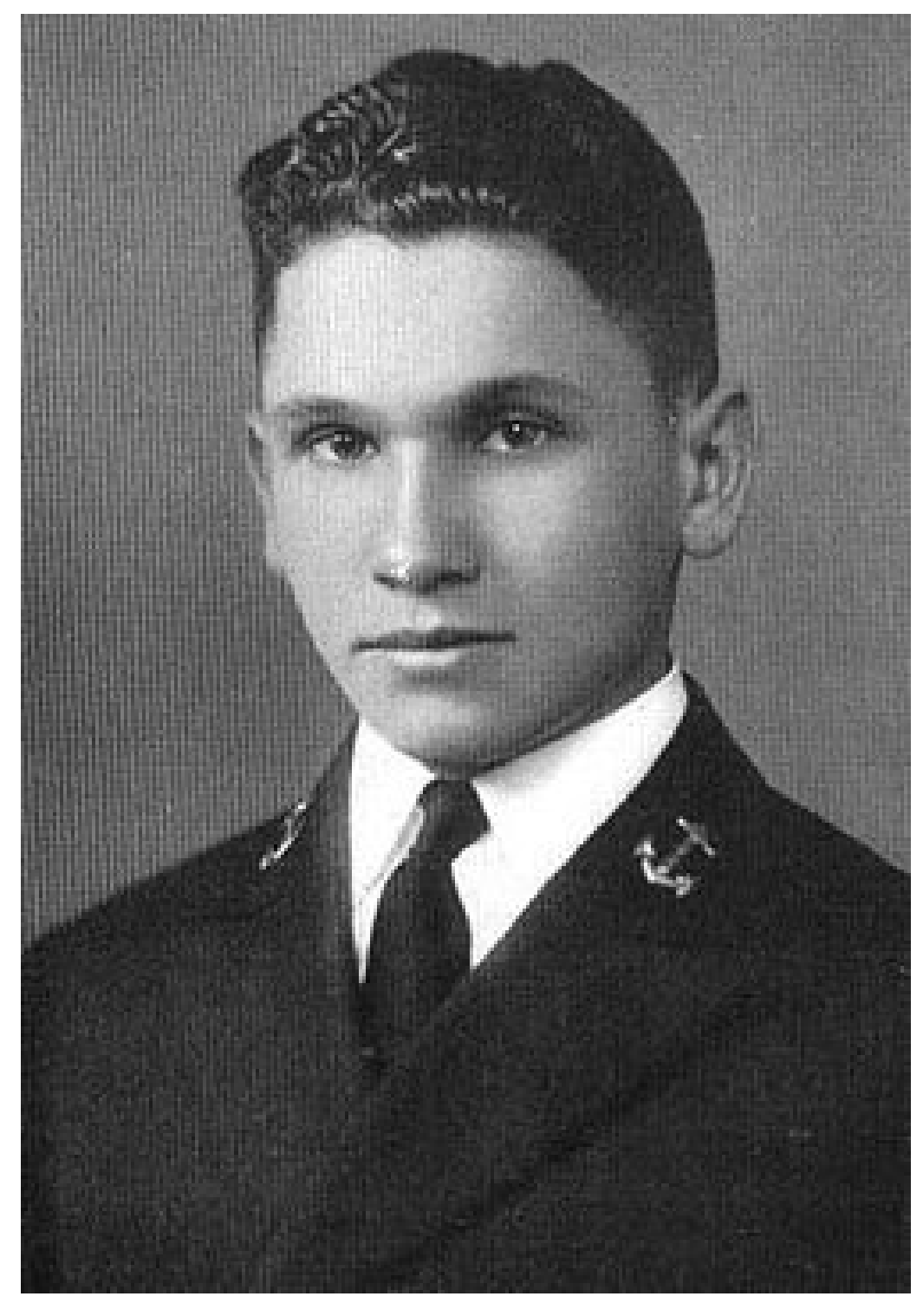

Во время второй мировой войны Вебер служил в военно-морском флоте США (фотография справа), стал капитаном корабля – охотника за подводными лодками. Он изучил электронику и в 1945 году возглавил отдел разработки средств радиоэлектронного подавления. В 1948 году он вышел в отставку и стал профессором электротехники. На эту высокооплачиваемую дольжность его приняли с условием, что он быстро защитит диссертацию. Поэтому он подошел к известному физику Георгию Гамову и спросил, не знает ли тот такую проблему в микроволновой спектроскопии, которая бы годилась для диссертации. Гамов ответил негатовно (см. книгу J. Levin, Black Hole Blues and Other Songs from Outer Space). Чтобы понять весь трагизм этого «нет», надо вспомнить, что как раз Гамов (совместно с Альфером и Германом) в 1948 году предсказал существование реликтового микроволнового излучения – главного свидетеля Большого Взрыва. Реликтовое излучение было открыто совершенно случайно в 1965 году американскими радиоастрономами Пензиасом и Вильсоном, за что они получили Нобелевскую премию в 1978 году. Нет сомнений, что если бы Гамов тогда ответил положительно, эта премия досталась бы Веберу и Гамову.

Но невезения Вебера на этом не закончились. Он был первым, кто предложил принцип работы мазера и лазера на престижной конференции в 1952 году, но его расчеты показывали такую низкую эффективность этих устройств, что он решил не развивать идею и не строить реальных прототипов (см. книгу M. Bertolotti, The History of the Laser). В результате, такие прототипы были построены другими людьми, и Нобелевскую премию в 1964 году получили Басов, Прохоров и Таунс.

Но самое большое невезение Вебера связано с гравитационными волнами. Стержень в знаменитом аргументе в защиту существования гравитационных волн хоть и в меньшей степени, но все равно испытывает влияние гравитационной волны. Если частота гравитационной волны совпадет с собственной частотой колебания стержня, то после прохождения гравитационной волны стержень начнет едва заметно вибрировать. Именно на основе этой идеи Вебер и решил построить свой детектор гравитационных волн. В своем стремлении Вебер был гениально изобретательным, цепким и дерзким. Он действительно построил детектор, который работал. Это был 1.2-тонный алюминиевый цилиндр длиной около 1.5 м и диаметром около 61 см, подвешенный в вакуумной камере на акустических фильтрах. Его резонансная частота при комнатной температуре была 1657 Гц. Система датчиков, способная преобразовывать механическое напряжение в электрическое, представляла собой серию пьезоэлектрических кварцевых кристаллов, установленных на ее поверхности вблизи центральной области. Детектор начал работать с хорошей чувствительностью и с шумовой изоляцией в январе 1965 года. Два года спустя [17] Вебер сообщил о наблюдении первых возможных сигналов гравитационных волн.

В 1968 году он снова сообщил о возможных событиях. На этот раз он использовал два детектора, разнесенные на 2 км друг от друга, и утверждал, что зафиксировал совпадающие события.

Наконец, в 1969 году Вебер заявил, что совпадения наблюдались на детекторах гравитационного излучения, расположенных на расстоянии около 1000 км друг от друга. Он утверждал, что вероятность случайности этих совпадений была невероятно мала. Эти наблюдения Вебера вызвали большой ажиотаж в научном мире. Он стал, возможно, самым знаменитым физиком того времени. Однако, величина и частота появления зарегистрированных им сигналов были намного выше ожидаемого.

Новаторская работа Вебера была решающей [17] для первоначального роста гравитационно-волнового научного сообщества. Около десяти групп попытались повторить его результаты. Несмотря на то, что никто не смог подтвердить правоту Вебера, основы экспериментального изучения гравитационных волн были надежно заложены.

Американский биофизик Беверли Рубик как-то сказала [18]: «Первопроходцев всегда можно узнать по стрелам в их спине. Слабые падают от стрел. Но великие, несмотря ни на что, продолжают идти вперед, по крайней мере с дюжиной стрел, застрявших в их спине... Цель состоит в том, чтобы продвигаться настолько вперед, чтобы стрелы больше не могли дотянуться до вас. И тогда, может быть, вы сможете начать вытаскивать некоторые из них». Увы, Вебер не смог продвинутся достаточно далеко, и стрелы недоброжелателей и завистников достигали его. После того, как его результаты не подтвердились, научное сообщество отвернулось от него. Его чуть не уволили из его родного университета штата Мэриленд. Социолог Гарри Коллинз вспоминает, как Вебер сказал ему с грустной улыбкой о своей второй жене Вирджинии Тримбл, которая тогда была молодым астрономом на двадцать три года младше его: «Когда я женился на ней, я был знаменит, а она нет. А теперь наши роли поменялись» (см. книгу J. Levin, Black Hole Blues and Other Songs from Outer Space).

Нобелевскую премию за открытие гравитационных волн в 2017 году, уже после смерти Вебера (он умер в 2000 году), получили другие люди: Райнер Вайсс, Кип Торн и Барри Бэриш (на фотографии слева направо).

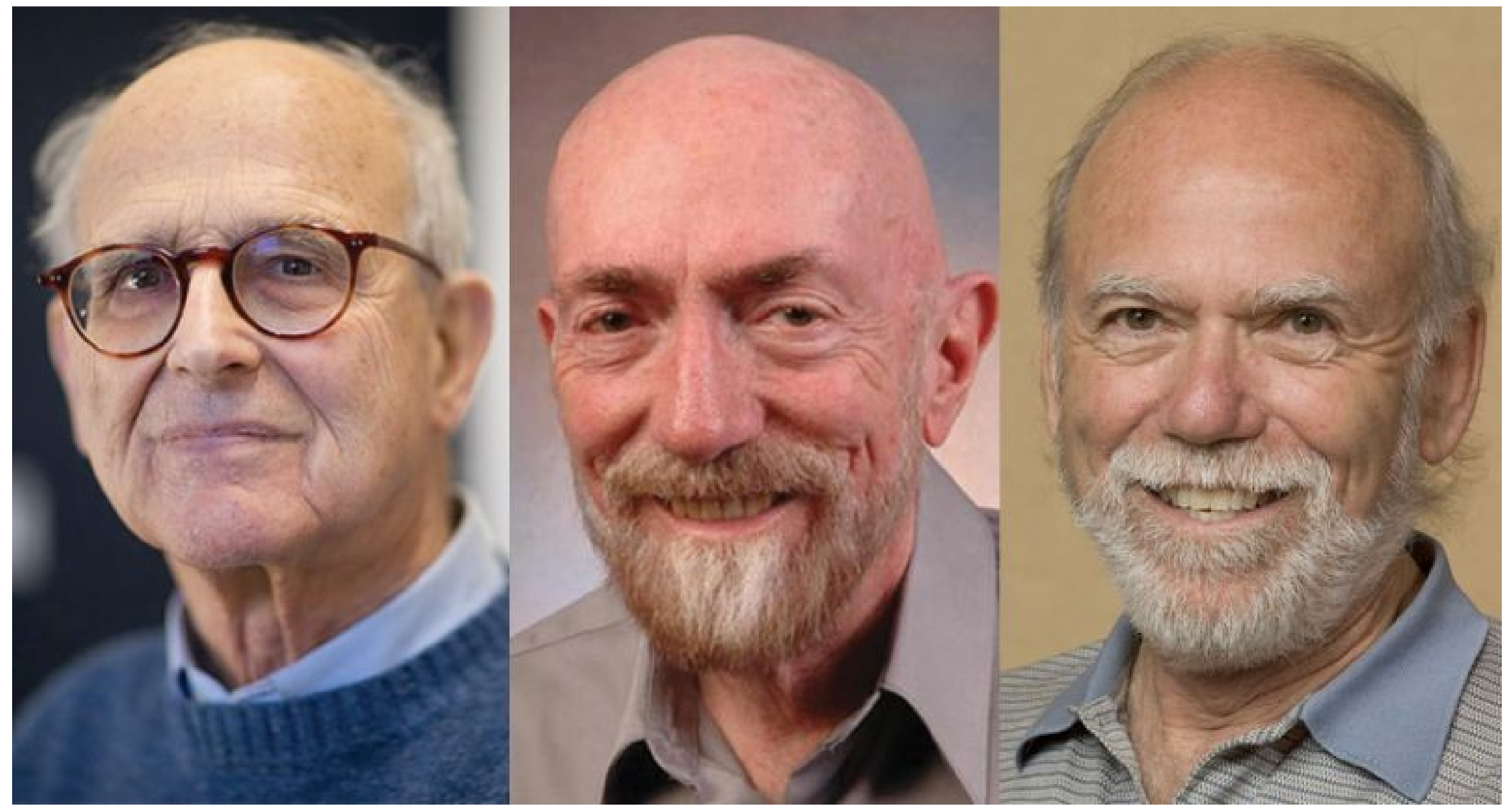

Райнер Вайсс не присутствовал на конференции в Чапел Хилле, но эта конференция все равно привела [16] к тому, хотя и менее прямолинейным способом, что Вайсс стал заниматься гравитационными волнами и, в конце концов, оказался у истоков лазерно-интерферометрической гравитационно-волновой обсерватории LIGO.

В 1964 году Вайссу предложили преподавательскую работу в аспирантуре Массачусетского технологического института по теории относительности. Стараясь как можно лучше объяснить предмет аспирантам, Вайсс задумался о том какие предсказания общей теории относительности действительно измеримы. Обнаружив работу Феликса Пирани 1956 года (одна из двух работ Пирани, представленных им на конференции в Чапел Хилле), Вайсс понял, что нашел ясный и убедительный способ ответить на этот вопрос.

Бусинки в мысленном эксперименте Фейнмана-Бонди под действием гравитационной волны смещаются сильнее, чем деформируется стержень. Поэтому Вайсс подумал, что можно достичь гораздо большей чувствительности, если в детекторе гравитационных волн использовать свободные массы. Эта идея в зачаточном виде присутствует уже в работе Пирани, когда тот рассматривал эффект прохождения гравитационной волны на группу свободных частиц.

В 1972 году в квартальном отчете о проделанной работе исследовательской лаборатории электроники Массачусетского технологического института появилась работа Вайсса, одна из самых важных неопубликованных работ в физике XX века [16], в которой он не только обосновал применение лазерной интерферометрии для детектирования гравитационных волн, но и дал подробный анализ многих источников шума, с которыми невероятно слабые сигналы от гравитационных волн должны будут конкурировать в процессе обработки данных с гравитационно-волнового детектора.

Заметим, что Вайсс был не первым, кто предложил использование лазерной интерферометрии для детектирования гравитационных волн. Еще в 1962 году советские физики Владислав Пустовойт (на фотографии слева) и Михаил Герценштейн (на фотографии справа) предложили эту идею. Но, к сожалению, в Советском Союзе не было предпринято никаких практических шагов для реализации данного подхода к детектированию гравитационных волн. Идея использования лазерной интерферометрии для детектирования гравитационных волн была в конце концов воскрешена в Советском Союзе в 1966 году Владимиром Брагинским, но затем снова ушла в небытие. Брагинский (на фотографии в середине он показан вместе с Кипом Торном) в последствии основал Московское отделение проекта LIGO и играл важную роль в коллаборации, предложив несколько ключевых идей.

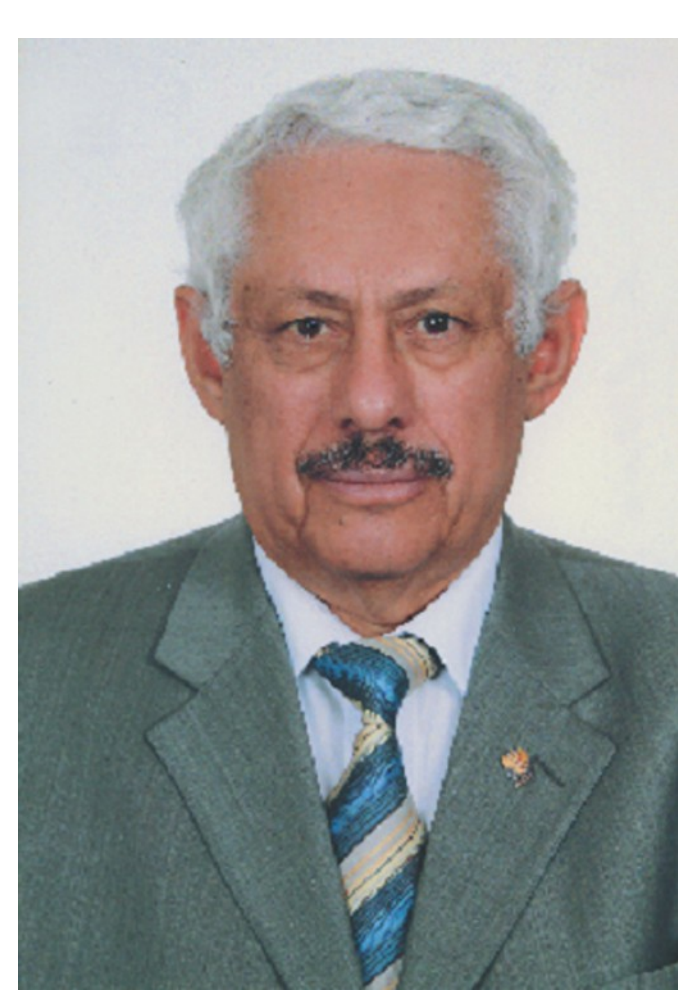

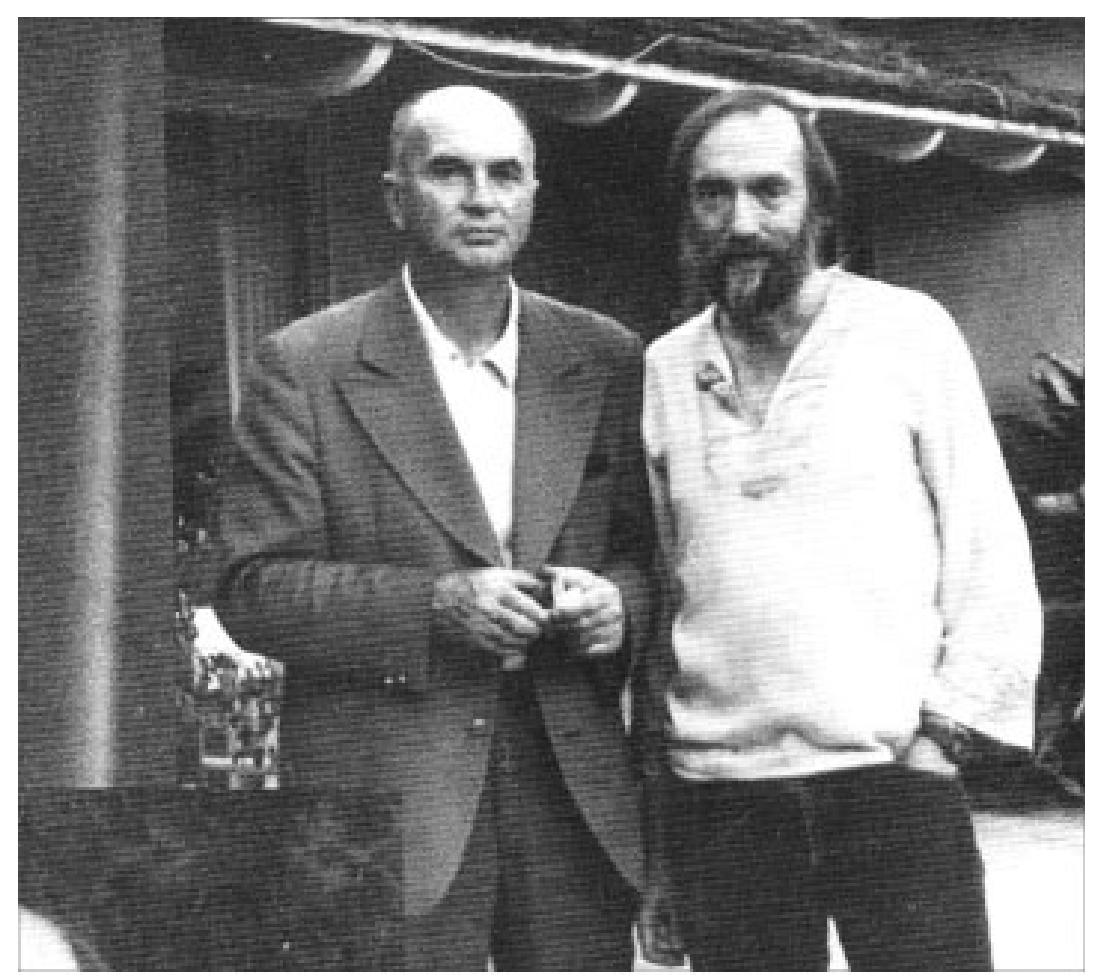

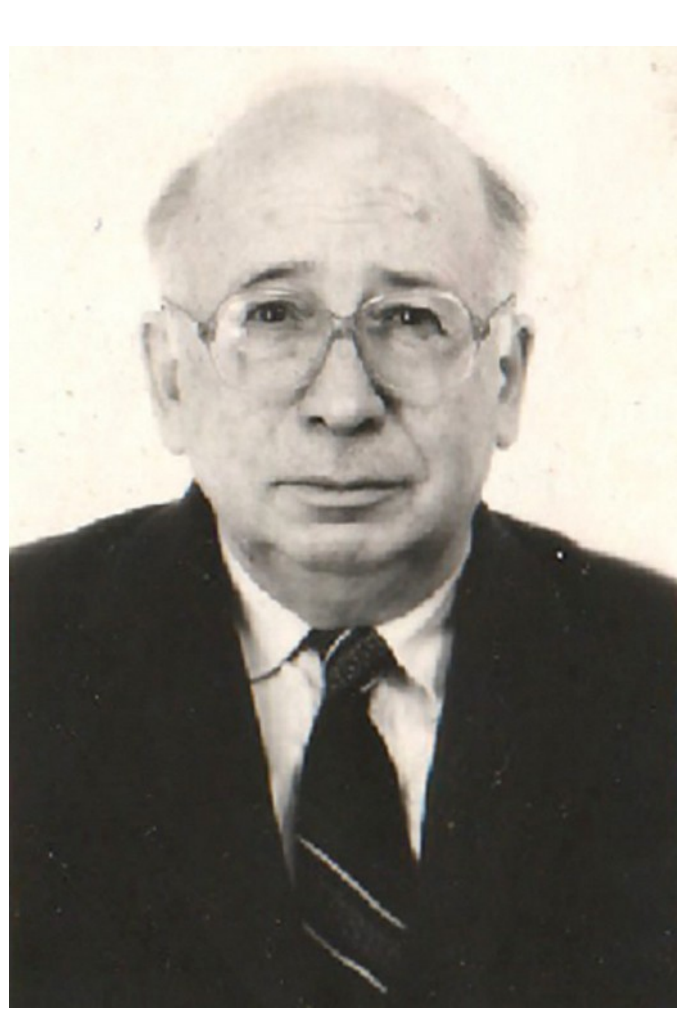

В 1970 году идея уже висела воздухе. В своей работе 1972 года Вайсс пишет [19]: «Идея не нова. Она появилась в мысленном эксперименте Пирани об измеримых свойствах тензора Римана. Однако, идея, что с появлением лазеров стало реальным обнаружение гравитационных волн с помощью этой техники [интерферометрии], выросла из студенческого семинара, который я проводил в Массачусетском технологическом институте несколько лет назад, и была независимо обнаружена доктором Филипом Чепменом из NASA» (позже Чепмен станет астронавтом).

В 1971 году бывший аспирант Вебера, Роберт Форвард, тоже представил конструкцию первого прототипа интерферометра (который он назвал «Лазером-преобразователем») и даже построил действующий 8,5-метровый интерферометр. В своей публикации Форвард утверждает [19], что идея интерферометра восходит к Веберу, и что тот сообщил ему об этой идее в телефонном разговоре 14 сентября 1964 года.

Фактически эта тройка: Форвард, Вайсс и Чепмен (на фотографиях они слева направо) - и была движущей силой [20] первых серьезных работ по интерферометрическому детектированию гравитационных волн.

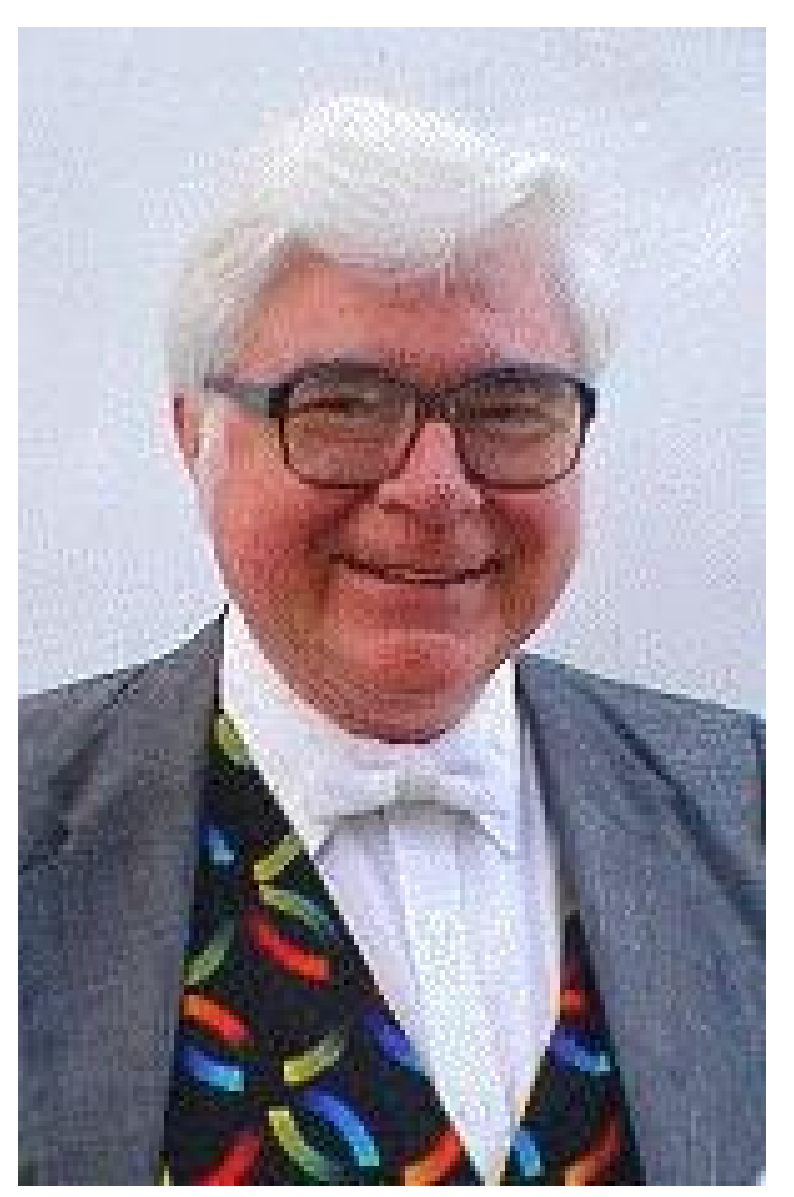 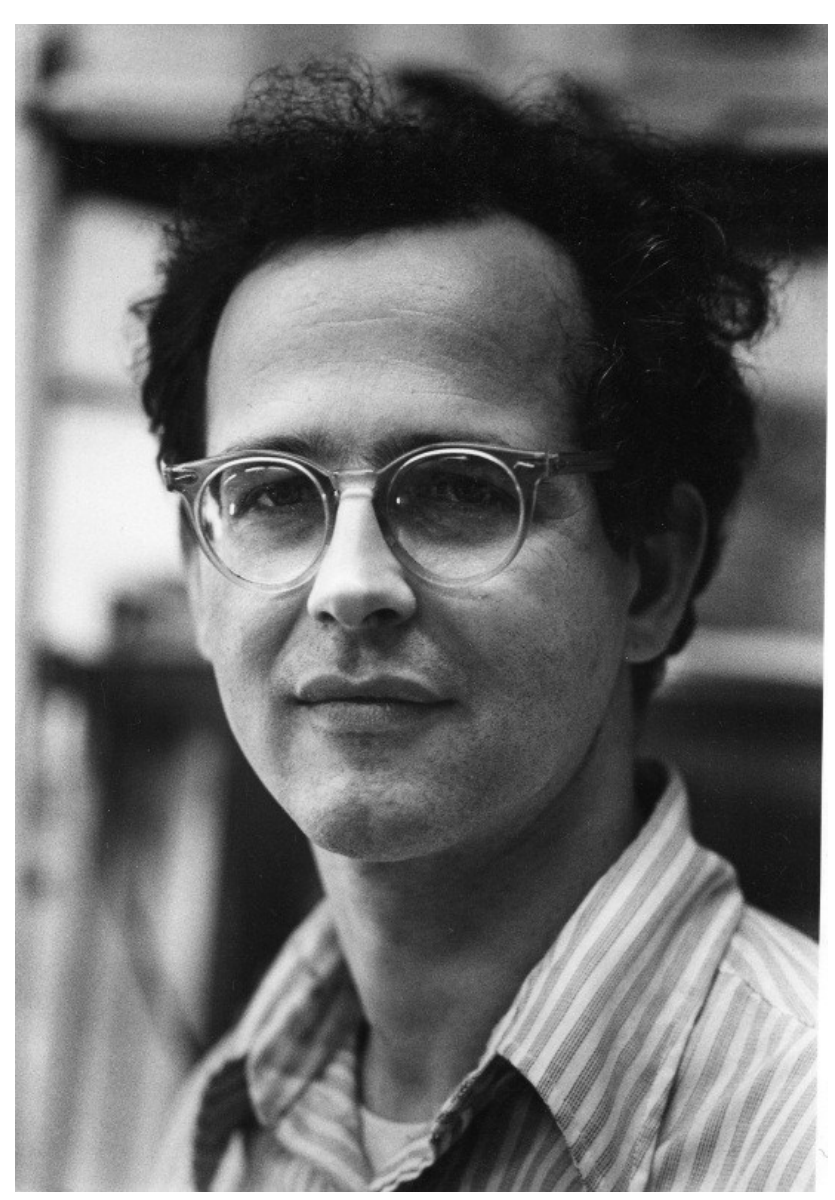 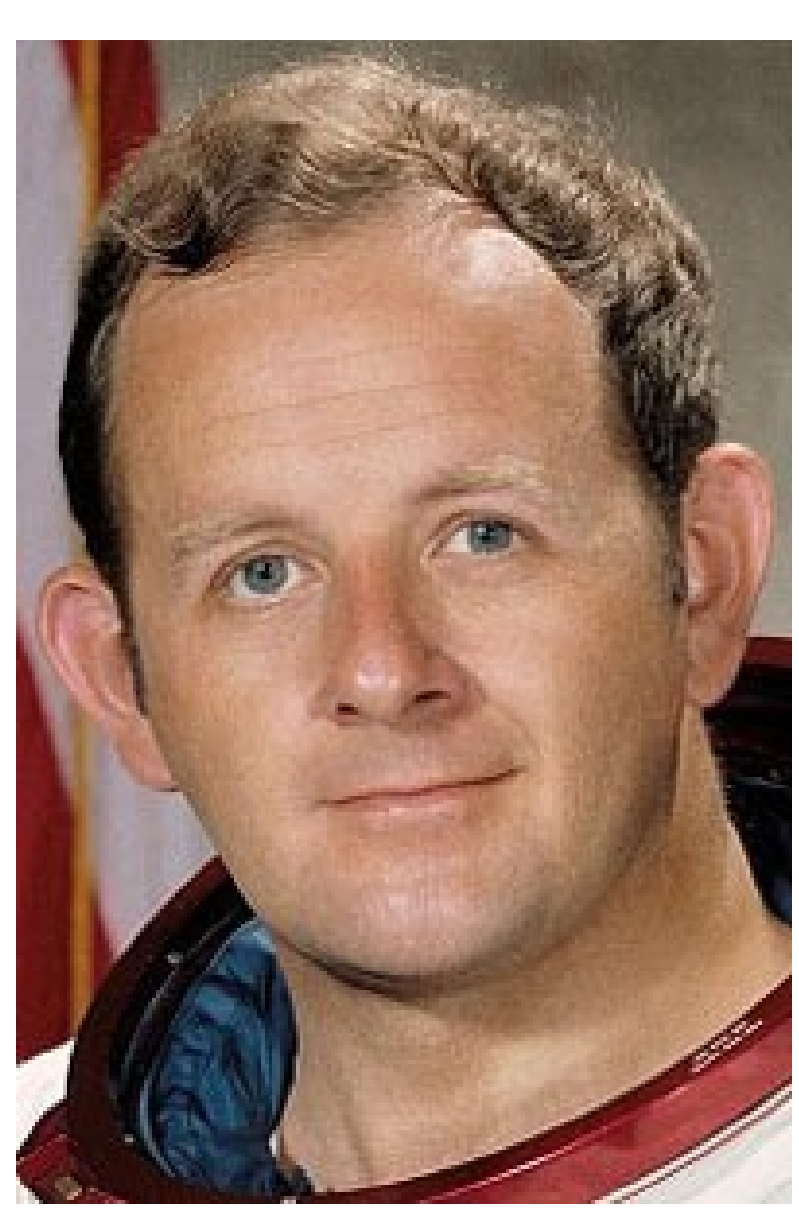

Вайсс строил свой 1,5-метровый прототип интерферометра на деньги военного ведомства. Но вскоре в США приняли закон, запрещающий вооруженным силам страны финансировать проекты, которые не были строго военными. Финансирование работ Вайсса было внезапно приостановлено, что вынудило его искать поддержку у других государственных и частных агентств США. Вайсс представил новый проект модернизации своего устройства (что предусматривало постройку 9-и метрового прототипа интерферометра) в Национальный Научный Фонд (NFS) США.

NFS попросил Питера Кафку из Института Макса Планка в Мюнхене рецензировать проект. Кафка был теоретиком и поэтому обратился за советом к своим друзьям-экспериментаторам, которые тогда совместно с итальянцами разрабатывали улучшенный вариант детектора Вебера.

К удивлению Кафки, местная экспериментальная группа пришла в восторг от проекта Вайсса и решила создать собственный прототип под руководством Хайнца Биллинга.

Немцы обратились за советом к Вайссу, который в конце концов прислал им своего студента Дэвида Шумейкера, имевшего опыта работы с 1,5-метровым прототипом Вайсса. Шумейкер позже помог построить немецкий 3-метровый прототип, а затем и 30-метровый интерферометр. На этом интерферометре отрабатывались методы подавления шумов, позже примененные в проекте LIGO. Со слов Вайсса [19], «группа Макса Планка действительно сделала большую часть важной предварительной работы. Они придумали много того, что я бы назвал практическими идеями, как сделать эту вещь [гравитационные интерферометры] намного лучше».

Немецкая группа несколько раз подавала заявку на финансирование большого проекта, но поддержку не получала. Наконец, они объединились с британской группой, основанной Рональдом Древером из университета Глазго, и подали заявку на финансирование 600-метрового интерферометра GEO-600. На этот раз их поддержали, и в сентябре 1995 года началось сооружение GEO-600. Помимо того, что GEO-600 был превосходной обсерваторией, он служил лабораторией [19] для разработки и испытания технологий, которые затем использовались в других детекторах гравитационных волн во всем мире.

Древер (на фотографии слева) не участвовал в постройке GEO-600, так как к этому времени он уже переехал в Калтех по приглашению Кипа Торна и возглавил исследования по детектированию гравитационных волн там (изначально Торн хотел пригласить Владимира Брагинского [20] возглавить мощную экспериментальную группу Калтеха по гравитационным волнам, но холодная война помешала этим планам). Древеру принадлежит идея многократного использования света внутри интерферометра с помощью оптического резонатора, что многократно снизило требуемую мощность лазера. Важный вклад в оживление исследований по общей теории относительности в Германии внес Бернард Шутц [21] (на фотографии справа). Он также сыграл большую роль в развитии плодотворного сотрудничества между теоретиками и экспериментаторами в астрофизике по всему миру, особенно в области обнаружения гравитационных волн.

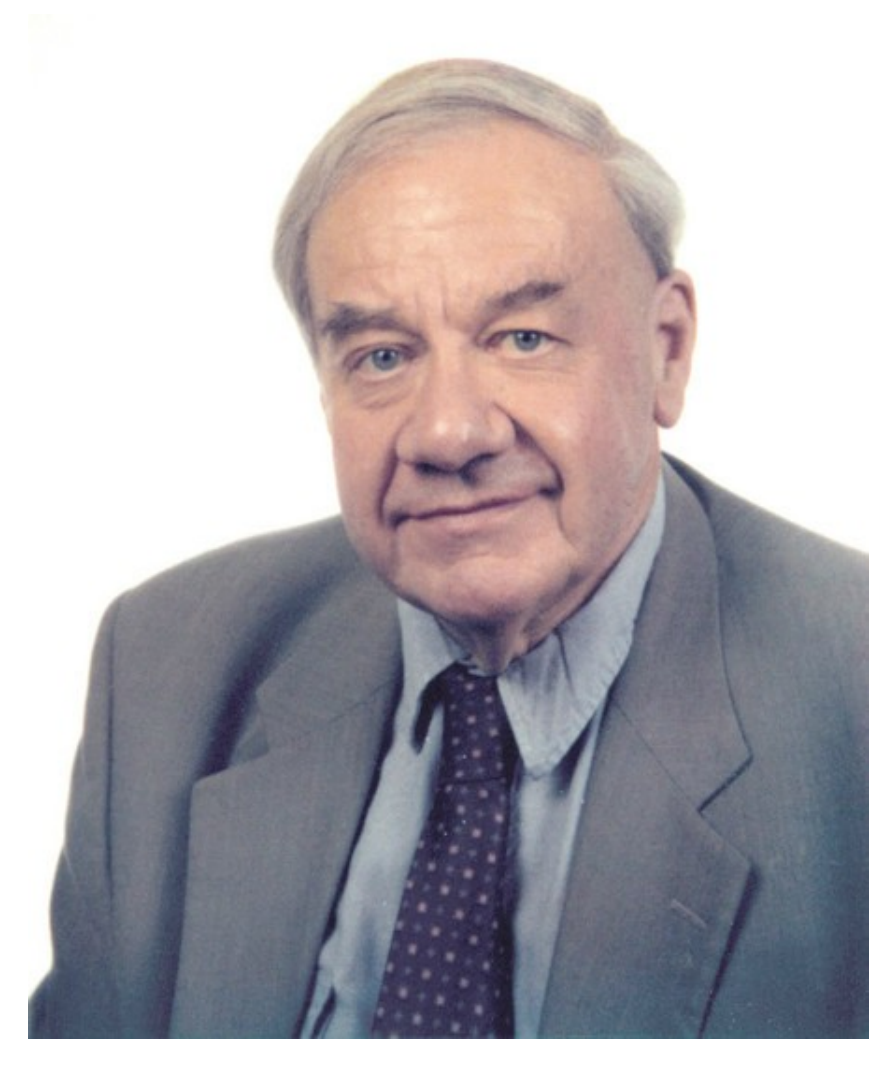

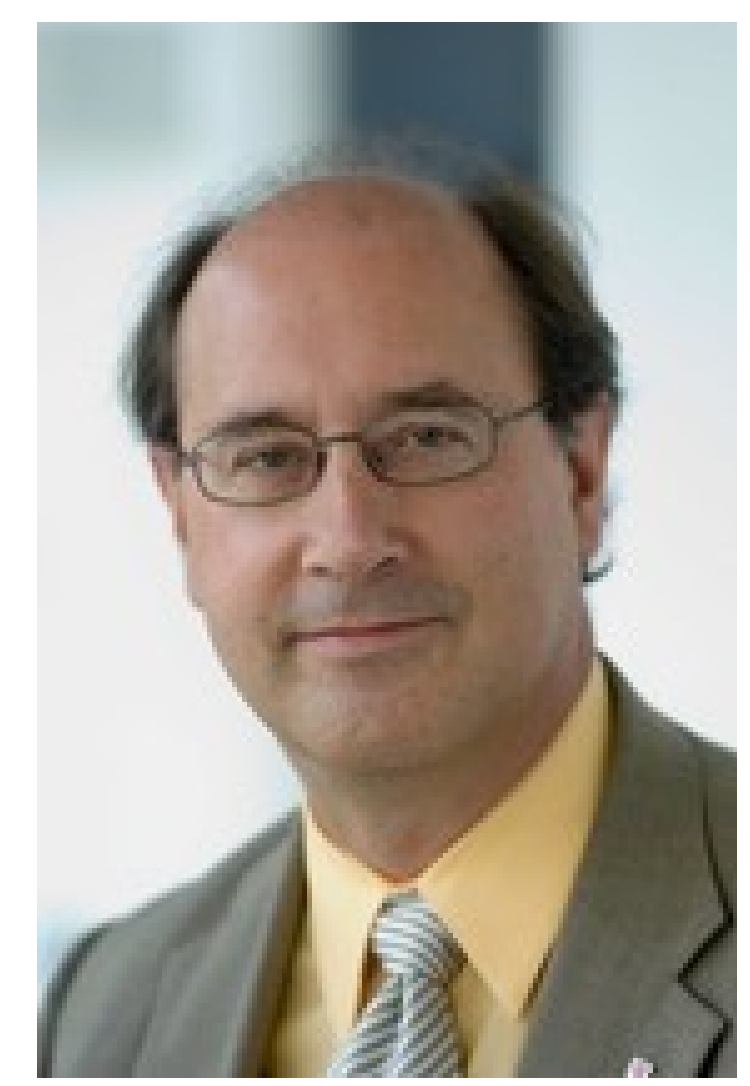

Кип Торн уже имел очень сильную теоретическую группу по общей теории относительности в Калтехе когда он стал задумываться о создании экспериментальной группы для изучения гравитационного излучения. Летом 1975 года судьба свела его с Вайссом, который объяснил ему суть интерферометрического метода. После того, как Торн понял, что это такое и почему он может достичь необходимой чувствительности для обнаружения гравитационных волн, он загорелся идеей. Это был поворотный момент [22] для проекта LIGO.

Как мы уже отметили, Торн пригласил в Калтех Древера в 1978 году, который вскоре, в 1983 году, построил 40-метровый прототип интерферометра. Вайсс решил, что настало время искать финансирование Национального Научного Фонда для полномасштабного проекта. Перед тем, как проект был представлен на рассмотрение NSF, Вайсс встретился с Торном и Древером [19] на конференции в Италии. Там они обсудили, как они могут работать вместе. С самого начала было ясно, что Древер не хотел сотрудничать с Вайссом, и Торн должен был действовать в качестве посредника.

Из-за многочисленных трений между Древером и Вайссом в течение 1984 и 1985 годов проект LIGO много раз задерживался. Наконец в 1986 году Национальный Научный Фонд потребовал упразднения трио Торна, Древера и Вайсса в качестве руководства проекта. Вместо этого Рохус Вогт был назначен главным менеджером проекта (на фотографии слева направо: Торн, Древер и Вогт около 40-метрового прототипа в 1990 году).

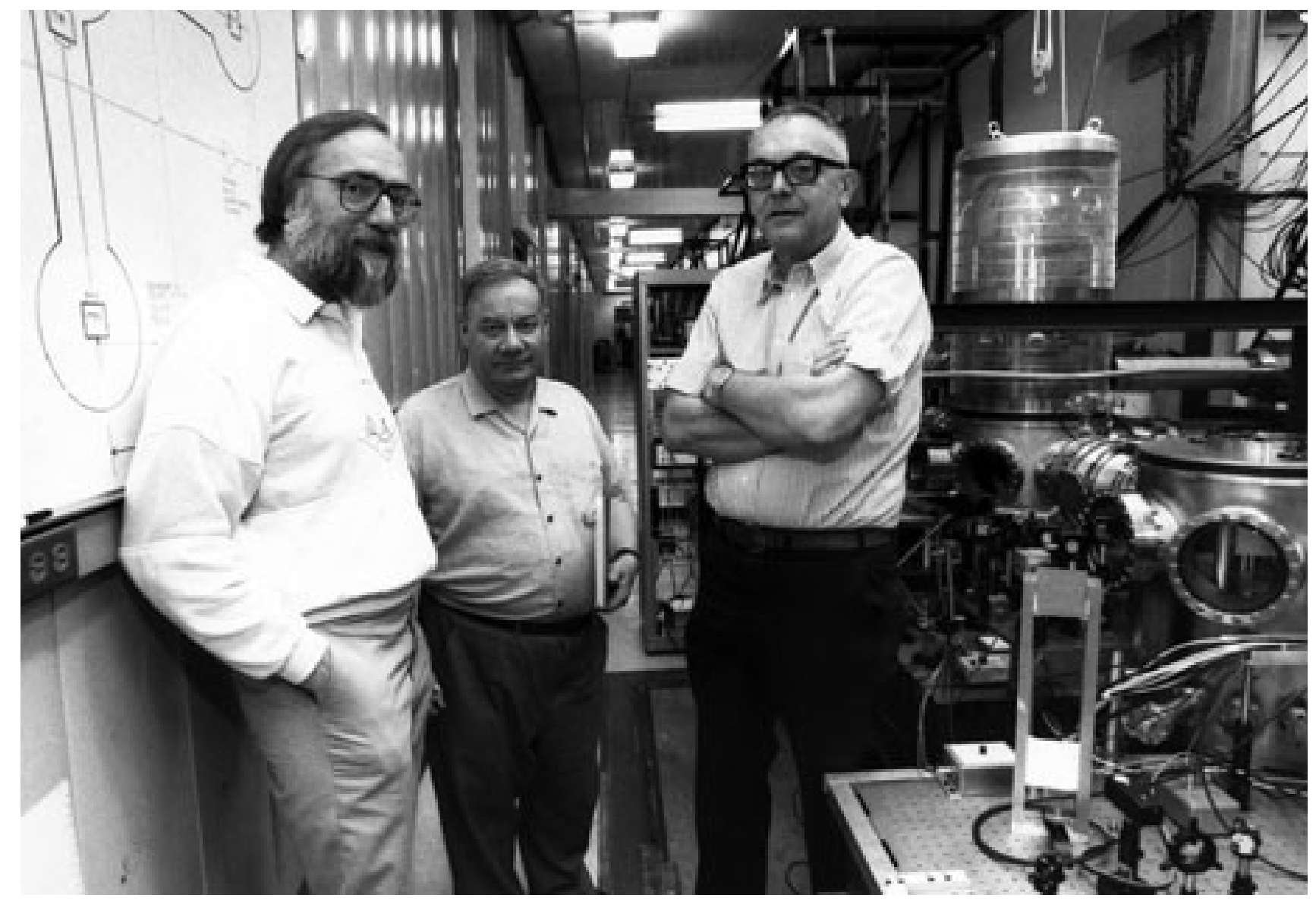

В 1988 году проект был окончательно профинансирован NSF [19]. Тем не менее прогресс проекта был медленным. В результате, Древер перестал участвовать проекте, а в 1994 году Вогт был заменен новым директором Барри Бэришем, имевшим опыт управления большими проектами в физике. Благодаря новому руководству проект получил хорошую финансовую поддержку и, в конце концов, закончился триумфом – в сентябре 2015 года была зарегистрирована первая гравитационная волна.

Другой полномасштабный проект детектирования гравитационных волн VIRGO был запущен в 1996 году французско-итальянской группой под руководством Аллена Брилле и  Адальберто Джиазотто (на фотографии они вместе). До этого оба ученые обратились к руководителям немецкого проекта GEO-600 из Института квантовой оптики им. Макса Планка в  Гархинге, надеясь сотрудничать с большим европейским детектором гравитационных волн. Но им отказали, сказав, что установление этого международного сотрудничества может задержать утверждение их уже почти поддержанного проекта. Поэтому они решили начать собственный параллельный проект, и только в 2007 году VIRGO и LIGO согласились начать сотрудничество по совместному поиску гравитационных волн.

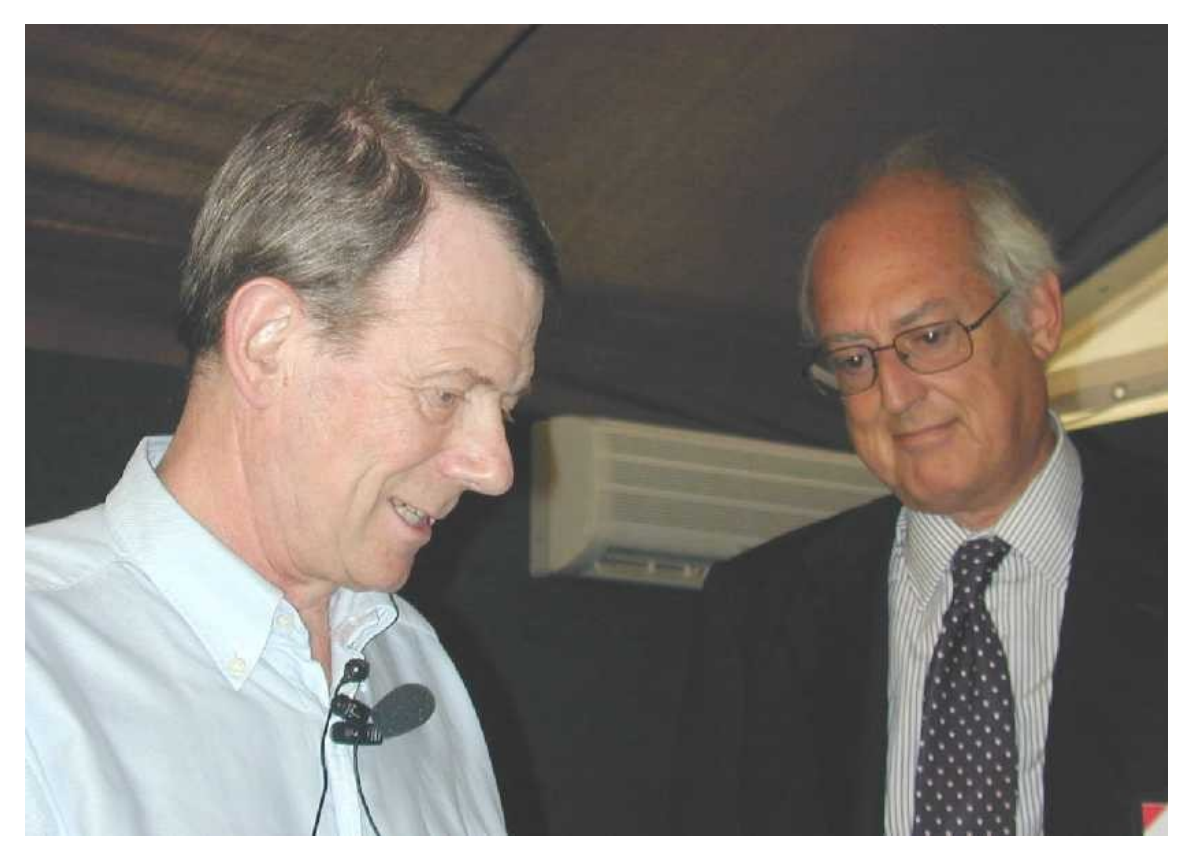

Почему отрытие гравитационных волн вызвало такой интерес среди не только научного собщества? Регистрация гравитационных волн дает не только еще один инструмент для изучения окружающего нас мира, но также приоткрывает завесу тайны над пространством-временем. Людям особенно хочется понять, что такое время, так как, как говорил немецкий философ Шпенглер, «со временем связана судьба». Специальная, а потом и общая теория относительности радикально изменили наши понятия о пространстве и времени. Лучше всего это передал сам Эйнштейн. Когда он незадолго до своей смерти узнал о кончине Мишеля Бессо, своего давнишнего друга, он сказал:  «Он ушел из этого странного мира чуть раньше меня. Это ничего не значит. Для нас, верующих физиков, разница между прошлым, настоящим и будущим – только иллюзия, за которую упрямо держатся». Как не держаться за эту иллюзию, если течение времени так ощутимо? Лучшим доказательством этого являются две следующие фотографии Эйнштейна: между ними время и судьба.

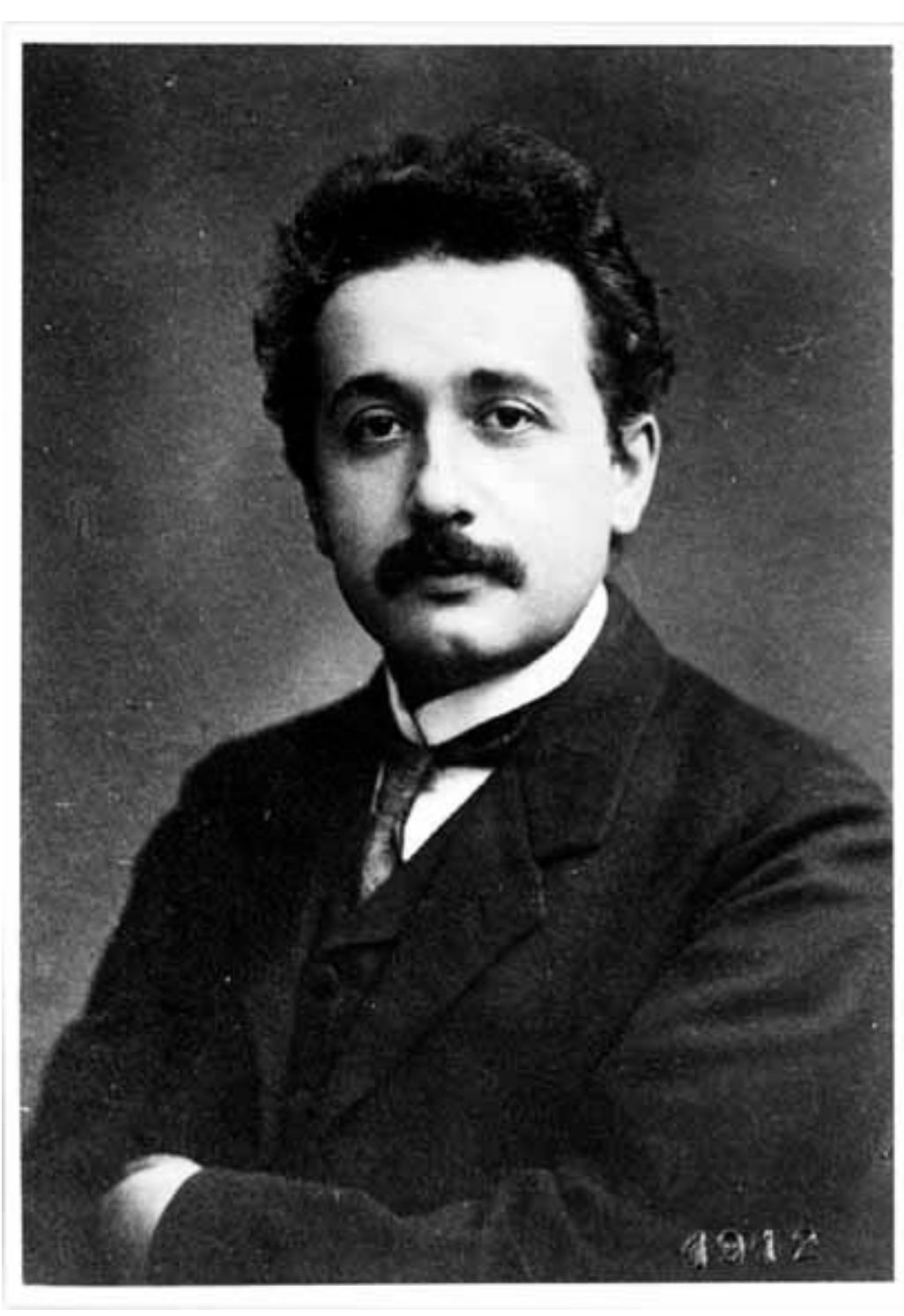

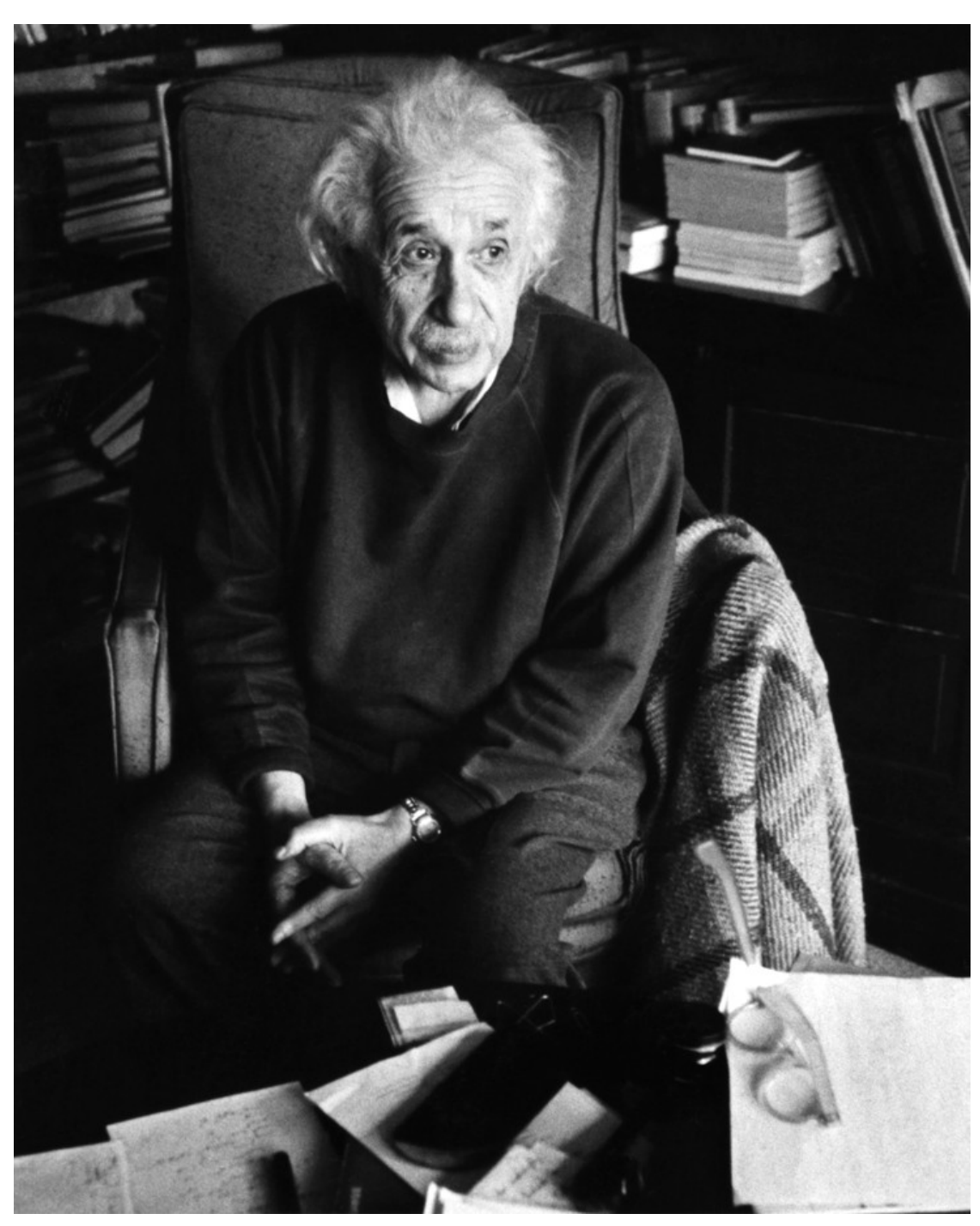

## Цитированная литература